\documentclass[twocolumn,english,showpacs,preprintnumbers,amsmath,amssymb,superscriptaddress]{revtex4}
\usepackage[T1]{fontenc}
\usepackage[latin9]{inputenc}
\usepackage{amsmath}
\usepackage{amssymb}
\usepackage{setspace}
\usepackage{esint}

\makeatletter
\@ifundefined{textcolor}{}
{%
 \definecolor{BLACK}{gray}{0}
 \definecolor{WHITE}{gray}{1}
 \definecolor{RED}{rgb}{1,0,0}
 \definecolor{GREEN}{rgb}{0,1,0}
 \definecolor{BLUE}{rgb}{0,0,1}
 \definecolor{CYAN}{cmyk}{1,0,0,0}
 \definecolor{MAGENTA}{cmyk}{0,1,0,0}
 \definecolor{YELLOW}{cmyk}{0,0,1,0}
 }

\@ifundefined{definecolor}{\usepackage{color}}{}
\usepackage{amsfonts}
\usepackage{fancyhdr}\usepackage{fancybox}\usepackage{ulem}

\usepackage{float}

\usepackage{amsthm}

\renewcommand{\Re}{{\rm Re}}
\renewcommand{\Im}{{\rm Im}}

\def\loweq@align#1#2{\lower.6ex\vbox{\baselineskip\z@skip\lineskip\z@
    \ialign{$\m@th#1\hfil##\hfil$\crcr#2\crcr=\crcr}}}
\def\lowsim@align#1#2{\lower.6ex\vbox{\baselineskip\z@skip\lineskip\z@
    \ialign{$\m@th#1\hfil##\hfil$\crcr#2\crcr\sim\crcr}}}
\def\geqq{\mathrel{\mathpalette\loweq@align >}}
\def\leqq{\mathrel{\mathpalette\loweq@align <}}
\def\grsim{\mathrel{\mathpalette\lowsim@align >}}
\def\lesssim{\mathrel{\mathpalette\lowsim@align <}}
\def\gsim{\mathrel{\mathpalette\lowsim@align >}}
\def\lsim{\mathrel{\mathpalette\lowsim@align <}}

\newcommand{\grless}{ {\, \raise-.24em\hbox{$<$} \hspace{-0.8em} \raise.31em\hbox{$>$}\, } }
\newcommand{\lessgr}{ {\, \raise-.24em\hbox{$>$} \hspace{-0.8em} \raise.31em\hbox{$<$}\, } }
\newfont{\bg}{cmr10 scaled\magstep4}                    
\newcommand{\bigzerou}{\smash{\lower1.7ex\hbox{\bg 0}}}

\newcommand{\crl}[1]{[-\infty,\infty]}

\usepackage{babel}

\usepackage{babel}

\makeatother

\usepackage{babel}
\begin{document}
\pagecolor{white}

\title{Time Dependent Quantum Mechanics}

\author{Peter G. Morrison}

\email{smallscience@hotmail.com}

\affiliation{Morrison Industrial Company, Sydney, Australia}

\date{August 1st, 2012}
\begin{abstract}
\noindent We present a systematic method for dealing with time dependent
quantum dynamics, based on the quantum brachistochrone and matrix
mechanics. We derive the explicit time dependence of the Hamiltonian
operator for a number of constrained finite systems from this formalism.
Once this has been achieved we go on to calculate the wavevector as
a function of time, in order to demonstrate the use of matrix methods
with respect to several concrete examples. Interesting results are
derived for elliptic curves and periodic orbits on higher dimensional
non-commutative geometries.


\end{abstract}


\maketitle

\section{Introduction}

Recent developments in quantum mechanical theory have focused on the
time-dependent dynamical problem, which proposes the use of Floquet
theory in order to derive the time evolution operators for the dependent
system. In our paper, we consider a new set of time-dependent quantum
mechanical operators which define the physics of finite systems. Standard
quantum mechanics postulates the existence of a Hamiltonian operator
$\widetilde{H}$ and an initial state $\left|\Psi(0)\right\rangle $
and seeks to extrapolate to find our final state $\left|\Psi(t)\right\rangle $,
via the von Neumann equation. This is an initial value problem; we
wish to find and calculate a boundary value problem that is consistent
with quantum physics. This has already been calculated in the papers
of Carlini et. al \cite{Carlini}. The principal methodology is to
construct a time-optimal Hamiltonian matrix which moves us from state-to-state
on the projective manifold, in least time. These paths are geodesic
flows; they define the surface of the state space in the same way
that the lines of longitude on the sphere define the globe itself.
In addressing this question, we are naturally led to the analysis
of explicitly time-dependent quantum systems. These dynamical systems
are generally periodic in nature for the examples we have been able
to calculate; this periodicity allows us to use Floquet theory in
order to find solutions for the problems in question. In considering
these types of general dynamical problems we find a number of time
dependent states with constant energies; wave-vector analysis allows
us to consider the geometry of the state space for higher-dimensional
realms where it is no longer possible to explicitly graph the paths
the state takes over time. Experiments have been conducted into these
areas with the particular aim of creating these exotic qutrit states,
as in \cite{Bogdanov}, \cite{Ranabir} and as such it is important
that we be able to predict best methods of state control within this
regime. Relevant references to the geometry of qutrits and SU(3) operators
may be found in \cite{Caves}.Our method differs from standard quantum
mechanics in that we do not assume the form of the Hamiltonian operator;
we derive it using the quantum brachistochrone hypothesis, being that
the physical path is the one of least time.

\section{Quantum Fermat Principle}

This section reproduces the arguments of \cite{Carlini} in deriving
the quantum brachistochrone equation from the Quantum Fermat Principle,
being the hypothesis that the state vector travels along a path of
least time on the complex projective manifold. These results are not
new but are included in order to outline the fundamentals of this
methodology. Firstly, let us consider the energy variance on the state
space. We may write this as:

\begin{equation}
(\Delta E)^{2}=\left\langle \psi\right|(\hat{H}-\left\langle \hat{H}\right\rangle \hat{\mathbf{1}})^{2}\left|\psi\right\rangle 
\end{equation}
We may define the metric of the projective state space: 
\begin{equation}
ds=\Delta E.dt
\end{equation}
The quantum state may be expanded in an orthonormal basis over a set
of probabilities:
\begin{equation}
\left|\psi\right\rangle =\sum_{j}\sqrt{p_{j}}e^{i\varphi_{j}}\left|j\right\rangle 
\end{equation}
which then define the state space metric:
\begin{equation}
\dfrac{1}{4}ds^{2}=1-\left|\left\langle \psi+d\psi|\psi\right\rangle \right|^{2}=\left\langle d\psi\left|(\hat{\mathbf{1}}-\hat{P})\right|d\psi\right\rangle 
\end{equation}
where $\hat{P}=\left|\psi\left\rangle \right\langle \psi\right|$,
the standard pure state projection operator. For the time-dependent
quantum control problem, the action principle must be recast in such
a way that we can apply it to find optimal transfers between states.
Our solution, where it exists, will be a time dependent Hamiltonian
that takes us time optimally from one state to another. We do not
rule out the existence of systems where there are a number of extremal
trajectories; however, our principal aim to search for global extrema
where the time is a minimum regardless of our particular choice of
initial and final states. Writing out the expression for the time
optimal quantum control action we have: 
\begin{equation}
S=\int_{t_{0}}^{t_{f}}(\mathcal{L}_{T}+\mathcal{L}_{S}+\mathcal{L}_{C})dt
\end{equation}
\begin{equation}
\mathcal{L}_{T}=\dfrac{\sqrt{\left\langle \dot{\psi}\left|(\hat{\mathbf{1}}-\hat{P})\right|\dot{\psi}\right\rangle }}{\Delta E}=\dfrac{1}{\Delta E}\dfrac{ds}{dt}
\end{equation}
\[
\mathcal{L}_{S}=i(\left\langle \phi\left|\right.\dot{\psi}\right\rangle -\left\langle \dot{\psi}\left|\right.\phi\right\rangle )-(\left\langle \phi\left|\hat{H}\right|\psi\right\rangle +\left\langle \psi\left|\hat{H}\right|\phi\right\rangle )
\]
\begin{equation}
\mathcal{L}_{C}=\sum_{j}\lambda_{j}f_{j}(\hat{H})
\end{equation}
These equations are gauge invariant; we therefore make a particular
choice of the gauge, being the reference frame in which the Hamiltonian
is traceless:
\begin{equation}
\hat{H}\mapsto\hat{H}-\dfrac{\hat{\mathbf{1}}_{N}}{N}Tr(\hat{H})=\tilde{H}
\end{equation}
\begin{equation}
Tr(\tilde{H})=0
\end{equation}
The first term is unity:
\begin{equation}
\mathcal{L}_{T}=\dfrac{\Delta E}{\Delta E}=1
\end{equation}
and therefore contributes a term $\int dt$ to the action, justifying
our claim of time optimality. This is the quantum analogue of Fermat's
principle of least time. Writing the Euler-Lagrange equations that
result from the variations of the action integral, we find the system
of equations:
\begin{equation}
i\left|\dot{\psi}\right\rangle =\tilde{H}\left|\psi\right\rangle 
\end{equation}
\begin{equation}
-\dfrac{i}{2}\dfrac{d}{dt}(\dfrac{\tilde{H}-\left\langle \tilde{H}\right\rangle \hat{\mathbf{1}})}{\Delta E^{2}})\left|\psi\right\rangle =i\left|\dot{\phi}\right\rangle -\tilde{H}\left|\phi\right\rangle 
\end{equation}
\begin{equation}
\dfrac{(\{\tilde{H},\hat{P}\}-2\left\langle \tilde{H}\right\rangle \hat{P})}{2\Delta E^{2}}+\left|\phi\left\rangle \right\langle \psi\right|+\left|\psi\left\rangle \right\langle \phi\right|=\sum_{j}\lambda_{j}\dfrac{\partial f_{j}}{\partial\tilde{H}}
\end{equation}
\begin{equation}
f_{j}(\tilde{H})=0\,\forall j
\end{equation}
Defining our constraint operator:
\begin{equation}
\hat{A}=\sum_{j}\lambda_{j}\dfrac{\partial f_{j}}{\partial\tilde{H}}=\dfrac{(\{\tilde{H},\hat{P}\}-2\left\langle \tilde{H}\right\rangle \hat{P})}{2\Delta E^{2}}+\left|\phi\left\rangle \right\langle \psi\right|+\left|\psi\left\rangle \right\langle \phi\right|
\end{equation}
Taking traces and expectation values of the numerator of the first
term in the constraint operator:
\begin{equation}
Tr(\{\tilde{H},\hat{P}\}-2\left\langle \tilde{H}\right\rangle \hat{P})=0
\end{equation}
\begin{equation}
<(\{\tilde{H},\hat{P}\}-2\left\langle \tilde{H}\right\rangle \hat{P})>=0
\end{equation}
We derive the fundamental result:
\begin{equation}
<\hat{A}>=Tr(\hat{A})=\left\langle \psi\left|\right.\phi\right\rangle +\left\langle \phi\left|\right.\psi\right\rangle 
\end{equation}
Computing the anticommutator of the A-operator and the projector we
obtain: 
\begin{equation}
\left\{ \hat{A},\hat{P}\right\} =\hat{A}+\left\langle \hat{A}\right\rangle \hat{P}
\end{equation}
We now define an associated operator to specify the boundary condition
equations:
\begin{equation}
\hat{G}=\hat{A}-\left\langle \hat{A}\right\rangle \hat{P}
\end{equation}
\begin{equation}
\left\{ \hat{G}(t),\hat{P}(t)\right\} =\hat{G}(t)
\end{equation}
\begin{equation}
Tr(\hat{G})=2<\hat{G}>=0
\end{equation}
We now perform some simple algebraic manipulations amongst our original
Euler-Lagrange equations. Left-multiplying (11) with $\left\langle \phi\right|$
and the conjugate of (12) with $\left|\psi\right\rangle $ we find:
\begin{equation}
i\left\langle \phi\right.\left|\dot{\psi}\right\rangle =\left\langle \phi\right|\tilde{H}\left|\psi\right\rangle 
\end{equation}
\begin{equation}
i\left\langle \dot{\phi}\right.\left|\psi\right\rangle =-\left\langle \phi\right|\tilde{H}\left|\psi\right\rangle -\dfrac{i}{2}\left\langle \psi\right|\dfrac{d}{dt}(\dfrac{\tilde{H}-\left\langle \tilde{H}\right\rangle \hat{\mathbf{1}})}{\Delta E^{2}})\left|\psi\right\rangle 
\end{equation}
Evaluating the time derivative of the expectation value of the Hamiltonian:
\begin{equation}
\dfrac{d}{dt}<\tilde{H}>=\left\langle \dot{\psi}\right|\tilde{H}\left|\psi\right\rangle +\left\langle \psi\right|\tilde{H}\left|\dot{\psi}\right\rangle +<\dfrac{d\tilde{H}}{dt}>
\end{equation}
\begin{equation}
\dfrac{d}{dt}<\tilde{H}>=<\dfrac{d\tilde{H}}{dt}>
\end{equation}
This relationship allows us to differentiate under the expectation
sign. We now calculate the time derivative of the overlap:
\begin{equation}
\dfrac{d}{dt}\left\langle \phi\left|\right.\psi\right\rangle =-\dfrac{1}{2}<\dfrac{d}{dt}(\dfrac{\tilde{H}-\left\langle \tilde{H}\right\rangle \hat{\mathbf{1}}}{\Delta E^{2}})>=0
\end{equation}
Therefore we obtain a constant of the motion given by:
\begin{equation}
\left\langle \phi\left|\right.\psi\right\rangle =const.=-\left\langle \psi\left|\right.\phi\right\rangle +<\hat{A}>
\end{equation}
A similar exercise from the variational equation for the Hamiltonian
yields:
\begin{equation}
\left|\phi\right\rangle =(\left\langle \psi\left|\right.\phi\right\rangle -\dfrac{(\tilde{H}-\left\langle \tilde{H}\right\rangle \hat{\mathbf{1}})}{2\Delta E^{2}}+\hat{G})\left|\psi\right\rangle 
\end{equation}
Directly differentiating this expression with respect to time and
substituting the constant of the motion into the equation we derive
the fundamental law of motion:
\begin{equation}
(i\dfrac{d\hat{G}}{dt}+[\hat{G},\tilde{H}])\left|\psi\right\rangle =0
\end{equation}
Some trivial algebra using Hermitian conjugates and right multiplication
proves that the G-operator and projection operator follow the Heisenberg
equation of motion:
\begin{equation}
i\dfrac{d\hat{G}}{dt}=[\tilde{H},\hat{G}]
\end{equation}
\begin{equation}
i\dfrac{d\hat{P}}{dt}=[\tilde{H},\hat{P}]
\end{equation}
We then use these expressions to evaluate the time derivative of the
A-operator by direct differentiation:
\begin{equation}
i\dfrac{d\hat{A}}{dt}=[\tilde{H},\hat{A}]+i\hat{P}\dfrac{d}{dt}<\hat{A}>
\end{equation}
\begin{equation}
\dfrac{d}{dt}<\hat{A}>=0
\end{equation}
\begin{equation}
i\dfrac{d\hat{A}}{dt}=[\tilde{H},\hat{A}]
\end{equation}
It is thus proved that the G-operator, A-operator and projection operator
all obey the Heisenberg equation of motion. We shall refer to this
important relationship as the Quantum Brachistochrone Equation. This
expression forms the foundation of applied time optimal quantum state
control and we shall use it repeatedly throughout this paper. As part
of our quantum control methodology we will consider an energetic constraint
as a Lagrange multiplier in the action functional. This restriction
holds the total energy used in the process to be some finite value,
and is given by:
\begin{equation}
f_{0}(\tilde{H})=Tr(\dfrac{\tilde{H}^{2}}{2})-k=0
\end{equation}
The other constraints which restrict the degrees of freedom for the
Hamiltonian of the system will often be linear in $\tilde{H}$: 
\begin{equation}
f_{j}(\tilde{H})=Tr(\tilde{H}\hat{g}_{j})
\end{equation}
where the $\hat{g}_{j}$'s are traceless Hermitian generators of the
unitary space. In this particular instance the Hamiltonian and A-operator
will obey the matrix equations:
\begin{equation}
\hat{A}=\lambda_{0}\tilde{H}+\tilde{F}
\end{equation}
\begin{equation}
Tr(\tilde{H}\tilde{F})=0
\end{equation}
\begin{equation}
\tilde{F}=\sum_{j}\lambda_{j}\hat{g}_{j}
\end{equation}
where the Hamiltonian and associated linear constraint are tracefree
Hermitian matrices. We take these results as fundamental for the time
optimal evolution of quantum states and may write the equations of
motion for our quantum dynamical system as:

\begin{singlespace}
\noindent \begin{flushleft}
\begin{equation}
i\frac{d}{dt}\left|\Psi(t)\right\rangle =\tilde{H}[t]\left|\Psi(t)\right\rangle 
\end{equation}
 
\begin{equation}
i\frac{d}{dt}(\tilde{H}[t]+\tilde{F}[t])=\tilde{H}[t]\tilde{F}[t]-\tilde{F}[t]\tilde{H}[t]
\end{equation}
 
\begin{equation}
Tr(\tilde{H}[t]\tilde{F}[t])=0
\end{equation}
 
\begin{equation}
Tr(\tilde{H}^{2}[t]/2)=\mathrm{constant}
\end{equation}
 
\begin{equation}
\{\hat{G}(t),\hat{P}(t)\}=\hat{G}(t)
\end{equation}

\par\end{flushleft}
\end{singlespace}

\paragraph*{\textmd{\textup{where $\hat{G}=(\tilde{H}+\tilde{F})-\left\langle \tilde{H}+\tilde{F}\right\rangle _{\psi}$,
and the derivative indicates an explicit differentiation of the matrices
with respect to the time parameter. The constraint, boundary condition
and Hamiltonian are all Hermitian matrices, so they will evolve unitarily
via the Heisenberg equation. When we use co-ordinate transformations,
we must use partial derivatives, but the distinction is made clear.
Our formalism is explicitly time dependent, in that the matrices derived
will be (in general) functions of the time parameter. These are more
difficult dynamic problems to solve than the time invariant case,
as the matrices within the differential equations to be solved are
functions of time instead of constants. }}}

\section{Isometry Groups Of Differential Operators And Matrices}

In this paper, often we will use isometric transformations to move
between various different equivalent physical pictures. Mainly we
confine ourselves to the interaction picture or the co-rotating reference
frame of Floquet; as such it is quite important that we have a concrete
method in place for understanding the meaning and application of isometry
and similarity transforms. Firstly, consider the coupled matrix differential
equations:

\begin{equation}
i\dfrac{d}{dt}\left|\chi(t)\right\rangle =(\hat{U}^{-1}\tilde{H}\hat{U})\left|\chi(t)\right\rangle 
\end{equation}

\begin{equation}
i\dfrac{d\hat{U}}{dt}=(\sum_{k}\lambda_{k}(t)\hat{P}_{k})\hat{U}(t,0)
\end{equation}
We may then write the solution as: 
\begin{equation}
\left|\chi(t)\right\rangle =\hat{Q}(t,0)\left|\chi(0)\right\rangle 
\end{equation}
\begin{equation}
\hat{Q}(t,0)=exp(-i\int_{0}^{t}\hat{U}^{-1}(s,0)\tilde{H}(s)\hat{U}(s,0)ds)
\end{equation}
\begin{equation}
\hat{U}(s,0)=exp(-i\int_{0}^{s}\sum_{k}\lambda_{k}(s')\hat{P}_{k}ds')
\end{equation}
Another relevant picture we may use is the Floquet co-rotating reference
frame. In this case, we have dynamical equations of motion given by:
\begin{equation}
i\dfrac{d}{dt}\left|\varphi(t)\right\rangle =(\dfrac{\partial\hat{S}}{\partial t}+e^{-i\hat{S}}\tilde{H}(t)e^{+i\hat{S}})\left|\varphi(t)\right\rangle 
\end{equation}
which has a solution that can be written in closed form as:
\begin{equation}
\left|\varphi(t)\right\rangle =\hat{K}(t,0)\left|\varphi(0)\right\rangle 
\end{equation}
\begin{equation}
\hat{K}(t,0)=exp(-i\int_{0}^{t}\hat{W}^{-1}(s,0)\tilde{H}(s)\hat{W}(s,0)ds)
\end{equation}
\begin{equation}
\hat{W}(s,0)=exp(-i\int_{0}^{s}\dfrac{\partial\hat{S}}{\partial\tau}d\tau)
\end{equation}

\section{Time Optimal Control For SU(2)}

For our first example, we consider the simple SU(2) system. In this
case, as the state space is equivalent to a double cover of the Bloch
sphere, we expect to recover great circles as our solutions. Our fundamental
relations are:

\begin{equation}
Tr(\dfrac{\tilde{H}^{2}}{2})-k=0
\end{equation}
\begin{equation}
Tr(\tilde{H}\hat{\sigma}_{z})=0
\end{equation}
\begin{equation}
Tr(\tilde{H})=0
\end{equation}
We may write the expression for the Hamiltonian as the ansatz: 
\begin{equation}
\tilde{H}=\left[\begin{array}{cc}
\alpha & \varepsilon\\
\varepsilon^{*} & -\alpha
\end{array}\right]
\end{equation}
Evaluating the trace of the Hamiltonian multiplied with the sigma-z
matrix we find directly that $\alpha=0$. Our time dependent Hamiltonian
operator is then:
\begin{equation}
\tilde{H}=\left[\begin{array}{cc}
0 & \varepsilon\\
\varepsilon^{*} & 0
\end{array}\right]
\end{equation}
Computing the quantum brachistochrone and calculating the right hand
side of the expression via ordinary matrix multiplication: 
\begin{equation}
i\dfrac{d}{dt}(\tilde{H}+\tilde{F})=\tilde{H}\tilde{F}-\tilde{F}\tilde{H}
\end{equation}
we find the explicitly time dependent matrix which defines the optimal
control fields to be given by:
\begin{equation}
i\dfrac{d}{dt}\left[\begin{array}{cc}
\Omega & \varepsilon\\
\varepsilon^{*} & -\Omega
\end{array}\right]=2\Omega\left[\begin{array}{cc}
0 & -\varepsilon\\
\varepsilon^{*} & 0
\end{array}\right]
\end{equation}
By inspection, $\Omega=\textrm{const.}$ We may write the differential
equations for the complex control fields in the form: 
\begin{equation}
i\dfrac{d}{dt}\left[\begin{array}{c}
\varepsilon(t)\\
\varepsilon^{*}(t)
\end{array}\right]=2\Omega\left[\begin{array}{cc}
-1 & 0\\
0 & 1
\end{array}\right]\left[\begin{array}{c}
\varepsilon(t)\\
\varepsilon^{*}(t)
\end{array}\right]
\end{equation}
\begin{equation}
\tilde{H}_{opt}(t)=\left[\begin{array}{cc}
0 & \varepsilon(0)e^{2i\Omega t}\\
\varepsilon^{*}(0)e^{-2i\Omega t} & 0
\end{array}\right]
\end{equation}
At this point the rotating wave approximation would usually be invoked,
but as the constraint is a constant of the motion, we may write:
\begin{equation}
(\tilde{H}_{opt}(t)+\Omega\hat{\sigma}_{z})\hat{U}(t,0)=\hat{U}(t,0)(\tilde{H}_{opt}(0)+\Omega\hat{\sigma}_{z})
\end{equation}
Using the Schrödinger equation, we find the dynamical equation for
the unitary:
\begin{equation}
i\dfrac{d\hat{U}}{dt}=\hat{U}(\tilde{H}_{opt}(0)+\Omega\hat{\sigma}_{z})-\Omega\hat{\sigma}_{z}\hat{U}
\end{equation}
This expression has an explicit solution:
\begin{equation}
\hat{U}(t,0)=exp(i\Omega\hat{\sigma}_{z}t)exp(-i(\tilde{H}_{opt}(0)+\Omega\hat{\sigma}_{z})t)
\end{equation}
Some simple algebra using the identity $exp(i\vec{n}.\mathbf{\hat{\sigma}})=\cos\theta\hat{\mathbf{1}}+i\vec{n}.\mathbf{\hat{\sigma}}\sin\theta$gives
the expansion for our unitary over the generators of SU(2) as:
\[
+\dfrac{(\varepsilon(0)-\varepsilon^{*}(0))}{2\Omega'}\sin(\Omega't)[\hat{\sigma}_{x}\sin(\Omega t)+\hat{\sigma}_{y}\cos(\Omega t)]
\]
 
\begin{equation}
+i\hat{\sigma}_{z}[\cos(\Omega't)\sin(\Omega t)-\dfrac{\Omega}{\Omega'}\cos(\Omega t)\sin(\Omega't)]
\end{equation}
where $\Omega'=\sqrt{k+\Omega^{2}}$. We require a unitary that satisfies
boundary conditions of the form:
\begin{equation}
\hat{U}(T,0)\hat{\sigma}_{x}\hat{U}^{\dagger}(T,0)=-\hat{\sigma}_{x}
\end{equation}
which gives us the quantisation condition $\Omega'T=m\pi/2$, $m\in\mathbb{N}$.
Choosing an initial state $\left|\Psi(0)\right\rangle =\dfrac{1}{\sqrt{2}}[1,1]^{T}$and
a final state $\left|\Psi(T)\right\rangle =\dfrac{1}{\sqrt{2}}[1,-1]^{T}$,
we find from the boundary conditions $(5)$ the equivalent conditions
on the control fields: 
\begin{equation}
\varepsilon(0)=-\varepsilon^{*}(0)
\end{equation}
\begin{equation}
\varepsilon(T)=-\varepsilon^{*}(T)
\end{equation}
We may rewrite this as: 
\begin{equation}
\sin(2\Omega T)=0
\end{equation}
which gives a second quantisation condition:
\begin{equation}
\Omega T=\dfrac{n\pi}{2},n\in\mathbb{N}
\end{equation}
Combining this with our first quantisation condition, we find the
period of the wave-vector to be:
\begin{equation}
T^{2}=\dfrac{\pi^{2}}{4k}(n^{2}-m^{2})
\end{equation}
The left hand side of this expression is a positive number, and therefore
$n>m$. The minimum time of operation is then:
\begin{equation}
T_{min}=\dfrac{\pi}{2\sqrt{k}}
\end{equation}
which is of the form of the Heisenberg energy-time uncertainty principle.
We may write the Hamiltonian matrix as:
\begin{equation}
\tilde{H}_{opt}(t)=\nu_{0}(\sin2\Omega t\hat{\sigma}_{x}+\cos2\Omega t\hat{\sigma}_{y})
\end{equation}
where $\nu_{0}^{2}=k$. Hence the best we can do in terms of energy
and time is: 
\begin{equation}
\nu_{0}T_{min}=\dfrac{\pi}{2}
\end{equation}
In summary, we have managed to derive the energy-time relation for
a revolving quantum state on SU(2); the arc that is described by our
unitary time evolution operator has the appropriate symmetry to be
considered a sinusoid convolved with a great circle.

\section{Unitary Matrices For SU(2)}

For completeness, we list a number of useful unitary matrices that
appear in the quaternionic geodesic calculation. These are related
in various ways through phase angles and transformations; we shall
not go into this here as the principal part of the calculation is
new work on SU(3). These matrices are:

\begin{equation}
\hat{U}_{1}=\dfrac{1}{\sqrt{2}}\left[\begin{array}{cc}
1 & -1\\
1 & 1
\end{array}\right]
\end{equation}
\begin{equation}
\hat{U}_{2}=\dfrac{1}{\sqrt{2}}\left[\begin{array}{cc}
e^{-i\theta} & -e^{-i\theta}\\
e^{i\theta} & e^{-i\theta}
\end{array}\right]
\end{equation}
\begin{equation}
\hat{U}_{3}=i\hat{U}_{1}\hat{U}_{2}(\chi)=\left[\begin{array}{cc}
\sin\chi & -i\cos\chi\\
i\cos\chi & -\sin\chi
\end{array}\right]
\end{equation}
\begin{equation}
\hat{U}_{4}=\left[\begin{array}{cc}
0 & 1\\
1 & 0
\end{array}\right]
\end{equation}
\begin{equation}
\hat{U}_{5}=\dfrac{1}{\sqrt{2}}\left[\begin{array}{cc}
e^{-i\alpha} & -e^{-i\alpha}\\
1 & 1
\end{array}\right]
\end{equation}
\begin{equation}
\hat{U}_{6}=exp(-i\vartheta\hat{\sigma}_{z})=e^{i\vartheta}\left[\begin{array}{cc}
1 & 0\\
0 & e^{-2i\vartheta}
\end{array}\right]
\end{equation}
\begin{equation}
\hat{U}_{j}\hat{U}_{j}^{\dagger}=\hat{U}_{j}^{\dagger}\hat{U}_{j}=\hat{\mathbf{1}}
\end{equation}
One may transform various Hamiltonians and density matrices to find
interesting relationships between different arrangements of constraints:
\begin{equation}
\hat{H}_{P}(\varsigma)=\hat{U}_{P}(\varsigma)\hat{H}(\varsigma)\hat{U}_{P}^{\dagger}(\varsigma)
\end{equation}
\begin{equation}
\dfrac{1}{2}\left[\begin{array}{cc}
1 & e^{-i\alpha}\\
1 & -e^{-i\alpha}
\end{array}\right]\left[\begin{array}{cc}
0 & e^{-i\alpha}\\
e^{+i\alpha} & 0
\end{array}\right]\left[\begin{array}{cc}
1 & 1\\
e^{i\alpha} & -e^{i\alpha}
\end{array}\right]=\left[\begin{array}{cc}
1 & 0\\
0 & -1
\end{array}\right]
\end{equation}

\section{Bilinear Quantum Algebra On SU(2)}

\noindent \begin{flushleft}
Writing the quantum brachistochrone in vector form for a single qubit
we find: 
\par\end{flushleft}

\begin{equation}
i\dfrac{d}{dt}(\left|h(t)\right\rangle +\left|f(t)\right\rangle )=\sum\left|j\right\rangle \left\langle h(t)\right|\hat{A}_{j}\left|f(t)\right\rangle 
\end{equation}

\paragraph*{\textmd{\textup{where the $\left|j\right\rangle $'s are the standard
basis set, and the matrices $\hat{A}_{j}$ have the property that
$\left\langle h(t)\right|\hat{A}_{j}\left|f(t)\right\rangle +\left\langle f(t)\right|\hat{A}_{j}\left|h(t)\right\rangle =0$.
Let us calculate the general SU(2) case as an application of the formula.
In this situation we have system Hamiltonian and matrix constraints
given by:}}}

\begin{equation}
\tilde{H}(t)=\sum_{i}h_{i}(t)\hat{\sigma}_{i\,}\mathrm{and\,}\tilde{F}(t)=\sum_{j}f_{j}(t)\hat{\sigma}_{j}
\end{equation}
where the Pauli matrices have matrix representation

\begin{equation}
\hat{\sigma}_{i}\in\left\{ \left[\begin{array}{cc}
1 & 0\\
0 & -1
\end{array}\right],\left[\begin{array}{cc}
0 & 1\\
1 & 0
\end{array}\right],\left[\begin{array}{cc}
0 & -i\\
i & 0
\end{array}\right]\right\} 
\end{equation}
 Our constraints state that the Hamiltonian and $\tilde{F}(t)$ are
orthogonal, in that their dot product is zero:

\begin{equation}
Tr(\tilde{H}(t)\tilde{F}(t))=\left\langle h(t)\right|\left.f(t)\right\rangle =\left\langle f(t)\right|\left.h(t)\right\rangle =0
\end{equation}

\noindent \begin{flushleft}
$i\dfrac{d}{dt}\left[\begin{array}{cc}
(h_{z}+f_{z}) & (\varepsilon_{h}+\varepsilon_{f})\\
(\varepsilon_{h}^{*}+\varepsilon_{f}^{*}) & -(h_{z}+f_{z})
\end{array}\right]$ 
\par\end{flushleft}

\begin{equation}
=\left[\begin{array}{cc}
(\varepsilon_{h}\varepsilon_{f}^{*}-\varepsilon_{f}\varepsilon_{h}^{*}) & +2(h_{z}\varepsilon_{f}-f_{z}\varepsilon_{h})\\
-2(h_{z}\varepsilon_{f}^{*}-f_{z}\varepsilon_{h}^{*}) & -(\varepsilon_{h}\varepsilon_{f}^{*}-\varepsilon_{f}\varepsilon_{h}^{*})
\end{array}\right]
\end{equation}

\noindent This may be written in the equivalent vector form as:

\begin{equation}
i\dfrac{d}{dt}\left(\left[\begin{array}{c}
\sqrt{2}h_{z}\\
\varepsilon_{h}\\
\varepsilon_{h}^{*}
\end{array}\right]+\left[\begin{array}{c}
\sqrt{2}f_{z}\\
\varepsilon_{f}\\
\varepsilon_{f}^{*}
\end{array}\right]\right)=\left[\begin{array}{c}
\sqrt{2}(\varepsilon_{h}\varepsilon_{f}^{*}-\varepsilon_{f}\varepsilon_{h}^{*})\\
2(h_{z}\varepsilon_{f}-f_{z}\varepsilon_{h})\\
-2(h_{z}\varepsilon_{f}^{*}-f_{z}\varepsilon_{h}^{*})
\end{array}\right]
\end{equation}

\paragraph*{\textmd{\textup{Our Hamiltonian and constraint vectors obey the relation}}}

\begin{equation}
\left\langle f(t)\right|\hat{A}_{j}\left|h(t)\right\rangle +\left\langle h(t)\right|\hat{A}_{j}\left|f(t)\right\rangle =0
\end{equation}

\paragraph*{\textmd{\textup{and hence we finally obtain the control matrices
for the vector control system on SU(2) as:}}}

\begin{equation}
\hat{A}_{1}=\sqrt{2}\left[\begin{array}{ccc}
0 & 0 & 0\\
0 & 1 & 0\\
0 & 0 & -1
\end{array}\right]
\end{equation}

\begin{equation}
\hat{A}_{2}=\sqrt{2}\left[\begin{array}{ccc}
0 & -1 & 0\\
0 & 0 & 0\\
1 & 0 & 0
\end{array}\right]
\end{equation}

\begin{equation}
\hat{A}_{3}=\sqrt{2}\left[\begin{array}{ccc}
0 & 0 & 1\\
-1 & 0 & 0\\
0 & 0 & 0
\end{array}\right]
\end{equation}

\noindent \begin{flushleft}
By explicit calculation one can show the identity $\left\langle h(t)\right|\hat{A}_{j}\left|f(t)\right\rangle +\left\langle f(t)\right|\hat{A}_{j}\left|h(t)\right\rangle =0$
holds true for all $j$; adjoint states are defined on the space $\left[\begin{array}{ccc}
\mathbb{R^{\mathrm{n-1}}}, & \mathbb{C}^{k*}, & \mathbb{C}^{k}\end{array}\right]$, being the complex conjugate and transposed row vector of the column
we started with. Let us now apply this formulation to a well-known
example, being the time-dependent Hamiltonian used by Carlini et.
al in {[}1{]}. The purpose of writing the quantum brachistochrone
in vector form is that it enables us to consider the physics directly,
which helps with understanding. In this case we have Hamiltonian and
constraint matrices as given by: 
\par\end{flushleft}

\begin{equation}
\tilde{H}(t)=\left[\begin{array}{cc}
0 & \varepsilon_{h}(t)\\
\varepsilon_{h}^{*} & 0
\end{array}\right];\tilde{F}(t)=\left[\begin{array}{cc}
f_{z}(t) & 0\\
0 & -f_{z}(t)
\end{array}\right]
\end{equation}

\noindent \begin{flushleft}
The vectors defined by these matrices may be written as: 
\par\end{flushleft}

\begin{equation}
\left|h(t)\right\rangle =\left[\begin{array}{c}
0\\
\varepsilon_{h}(t)\\
\varepsilon_{h}^{*}(t)
\end{array}\right];\left|f(t)\right\rangle =\left[\begin{array}{c}
\sqrt{2}f_{z}(t)\\
0\\
0
\end{array}\right]
\end{equation}

\noindent \begin{flushleft}
where we have used the factor of $\sqrt{2}$ to maintain the trace
product relation. These are multivectors in many respects, having
a real part and two complex components. The link between the matrix
form and the vector control equation is clear; we may use either technique
in order to calculate the time optimal Hamiltonian operator. 
\par\end{flushleft}

\noindent \begin{flushleft}
In this situation we may write the quantum brachistochrone in vector
form as: 
\par\end{flushleft}

\noindent $i\dfrac{d}{dt}\left[\begin{array}{c}
\sqrt{2}f_{z}(t)\\
\varepsilon_{h}(t)\\
\varepsilon_{h}^{*}(t)
\end{array}\right]=$

~

\begin{singlespace}
$\,\,\left[\begin{array}{c}
1\\
0\\
0
\end{array}\right]\left[\begin{array}{ccc}
\sqrt{2}f_{z}, & 0, & 0\end{array}\right]\sqrt{2}\left[\begin{array}{ccc}
0 & 0 & 0\\
0 & 1 & 0\\
0 & 0 & -1
\end{array}\right]\left[\begin{array}{c}
0\\
\varepsilon_{h}\\
\varepsilon_{h}^{*}
\end{array}\right]$

$+\left[\begin{array}{c}
0\\
1\\
0
\end{array}\right]\left[\begin{array}{ccc}
\sqrt{2}f_{z}, & 0, & 0\end{array}\right]\sqrt{2}\left[\begin{array}{ccc}
0 & -1 & 0\\
0 & 0 & 0\\
1 & 0 & 0
\end{array}\right]\left[\begin{array}{c}
0\\
\varepsilon_{h}\\
\varepsilon_{h}^{*}
\end{array}\right]$

$+\left[\begin{array}{c}
0\\
0\\
1
\end{array}\right]\left[\begin{array}{ccc}
\sqrt{2}f_{z}, & 0, & 0\end{array}\right]\sqrt{2}\left[\begin{array}{ccc}
0 & 0 & 1\\
-1 & 0 & 0\\
0 & 0 & 0
\end{array}\right]\left[\begin{array}{c}
0\\
\varepsilon_{h}\\
\varepsilon_{h}^{*}
\end{array}\right]$ 
\end{singlespace}

\begin{flushleft}
\begin{equation}
\therefore i\dfrac{d}{dt}\left[\begin{array}{c}
\sqrt{2}f_{z}(t)\\
\varepsilon_{h}(t)\\
\varepsilon_{h}^{*}(t)
\end{array}\right]=\left[\begin{array}{c}
0\\
-2f_{z}\varepsilon_{h}\\
+2f_{z}\varepsilon_{h}^{*}
\end{array}\right]
\end{equation}

\par\end{flushleft}

\noindent \begin{flushleft}
as required. This system may be rewritten in the compact form: 
\par\end{flushleft}

\begin{equation}
\left\langle h\right|\hat{D}\left|h\right\rangle =\mathrm{const.}=\lambda^{2}+|\varepsilon|^{2}
\end{equation}
 
\begin{equation}
\left\langle f\right|\hat{E}\left|h\right\rangle =\left\langle h\right|\hat{E}\left|f\right\rangle =0=2\lambda\Gamma+\varepsilon^{*}\Pi+\varepsilon\Pi^{*}
\end{equation}
 
\begin{equation}
\left\langle f\right|\hat{B}_{j}\left|h\right\rangle =-\left\langle h\right|\hat{B}_{j}\left|f\right\rangle 
\end{equation}
 
\begin{equation}
i\dfrac{d}{dt}(\left|h\right\rangle +\left|f\right\rangle )=\sum_{j}\left|j\right\rangle \left\langle h\right|\hat{B}_{j}\left|f\right\rangle 
\end{equation}
 
\begin{eqnarray}
\hat{B}_{1}=\left[\begin{array}{ccc}
0 & 0 & 0\\
0 & -1 & 0\\
0 & 0 & 1
\end{array}\right] &  & \hat{B}_{2}=2\left[\begin{array}{ccc}
0 & 1 & 0\\
0 & 0 & 0\\
-1 & 0 & 0
\end{array}\right]
\end{eqnarray}
 
\begin{equation}
\hat{B}_{3}=2\left[\begin{array}{ccc}
0 & 0 & -1\\
1 & 0 & 0\\
0 & 0 & 0
\end{array}\right]
\end{equation}
 
\begin{equation}
\hat{D}=\frac{1}{2}\left[\begin{array}{ccc}
1 & 0 & 0\\
0 & \Upsilon & 0\\
0 & 0 & (1-\Upsilon)
\end{array}\right]
\end{equation}
 
\begin{eqnarray}
 & \hat{E}=\frac{1}{2}\left[\begin{array}{ccc}
1 & 0 & 0\\
0 & 1/2 & 0\\
0 & 0 & 1/2
\end{array}\right]
\end{eqnarray}
 
\begin{equation}
\left|h\right\rangle =\left[\begin{array}{c}
\lambda\\
\varepsilon\\
\varepsilon^{*}
\end{array}\right],\left|f\right\rangle =\left[\begin{array}{c}
\Gamma\\
\Pi\\
\Pi^{*}
\end{array}\right]\in\left[\begin{array}{c}
\mathbb{R}\\
\mathbb{C^{\downharpoonleft}}
\end{array}\right]
\end{equation}
 
\begin{equation}
\left\langle h\right|=\left[\lambda,\varepsilon^{*},\varepsilon\right],\left\langle f\right|=\left[\Gamma,\Pi^{*},\Pi\right]\in\left[\mathbb{R},\mathbb{C^{\rightharpoondown}}\right]
\end{equation}
These types of mappings must preserve inner product relationships
between the Hamiltonian and constraint.

\section{SO(3) Matrix Analysis}

Consider a Hamiltonian matrix of the form:

\begin{equation}
\tilde{H}[t]=\vec{n}(t)\centerdot\hat{S}=\left[\begin{array}{ccc}
n_{z}(t) & 0 & \varepsilon(t)\\
0 & 0 & 0\\
\varepsilon^{*}(t) & 0 & -n_{z}(t)
\end{array}\right]
\end{equation}
We form a constraint matrix:

\begin{equation}
\tilde{F}[t]=\left[\begin{array}{ccc}
u & K_{1} & 0\\
K_{1}^{*} & -2u & K_{2}\\
0 & K_{2}^{*} & u
\end{array}\right]
\end{equation}
The quantum brachistochrone equation reads as:

\begin{equation}
i\frac{d}{dt}(\tilde{H}(t)+\tilde{F}(t))=\tilde{H}(t)\tilde{F}(t)-\tilde{F}(t)\tilde{H}(t)
\end{equation}
Evaluating this for our matrix system, we find the set of differential
equations:

\begin{equation}
\dot{u}=\dot{n}_{z}=\dot{\varepsilon}=\dot{\varepsilon}^{*}=0;\tilde{H}[t]=\tilde{H}[0]
\end{equation}
Writing our wave-vector in standard form we find the relations: 
\begin{equation}
c_{2}(t)=c_{2}(0)
\end{equation}

\begin{eqnarray}
i\frac{d}{dt}\left[\begin{array}{c}
c_{1}(t)\\
c_{3}(t)
\end{array}\right]=\left[\begin{array}{cc}
n_{z} & \varepsilon\\
\varepsilon^{*} & -n_{z}
\end{array}\right]\left[\begin{array}{c}
c_{1}(t)\\
c_{3}(t)
\end{array}\right]
\end{eqnarray}

\begin{equation}
Tr(\tilde{H}^{2}/2)=\mathrm{const.}=n_{z}^{2}+|\varepsilon|^{2}=R^{2}
\end{equation}
The propagator may be written in the compact form: 
\begin{equation}
\hat{U}(t,0)=\hat{\mathbf{1}}-i\frac{\mathrm{sin}Rt}{R}\tilde{H}[0]+(\mathrm{cos}Rt-1)\left[\begin{array}{ccc}
1 & 0 & 0\\
0 & 0 & 0\\
0 & 0 & 1
\end{array}\right]
\end{equation}
Choosing an initial state $\left|\psi(0)\right\rangle =\left[\begin{array}{ccc}
1 & 0 & 0\end{array}\right]^{T}$ we find the wave-vector:

\begin{equation}
\left|\psi(t)\right\rangle =\hat{U}(t,0)\left|\psi(0)\right\rangle =\left[\begin{array}{c}
\mathrm{cos}Rt-i\frac{n_{z}}{R}\mathrm{sin}Rt\\
0\\
-\frac{i}{R}\varepsilon^{*}\mathrm{sin}Rt
\end{array}\right]
\end{equation}
This is $2\pi-$periodic and satisfies $-\left|\psi(\frac{m\pi}{R})\right\rangle =\left|\psi(0)\right\rangle =\left|\psi(\frac{2m\pi}{R})\right\rangle $,
where m is an integer.

\section{Time Dependent Elliptic Hamiltonian on SU(3)}

\noindent In this section we shall consider three examples of SU(3)
Hamiltonians. The physical motivation for this is to explicitly calculate
brachistochrones for some model problems which are more general than
the SU(2) models considered in \cite{Carlini}. Boscain et. al has
examined various situations on SU(3) that have similar properties,
see for example \cite{Boscain}. Our first Hamiltonian matrix is given
by:

\noindent 
\begin{equation}
\tilde{H}(t)=\left[\begin{array}{ccc}
0 & \alpha(t) & 0\\
\alpha(t) & 0 & -i\beta(t)\\
0 & i\beta(t) & 0
\end{array}\right]
\end{equation}

\begin{singlespace}
\noindent with an associated constraint matrix 
\end{singlespace}

\begin{equation}
\tilde{F}(t)=\left[\begin{array}{ccc}
\omega_{1}(t) & -i\gamma(t) & \varepsilon(t)\\
i\gamma(t) & \omega_{2}(t) & \kappa(t)\\
\varepsilon^{*}(t) & \kappa(t) & \omega_{3}(t)
\end{array}\right];\sum_{n=1}^{3}\omega_{j}=0.
\end{equation}

\noindent The quantum brachistochrone equation in vector form then
reads as:

\begin{equation}
\frac{d}{dt}\left[\begin{array}{c}
\omega_{1}(t)\\
\omega_{2}(t)\\
\omega_{3}(t)
\end{array}\right]=2\left[\begin{array}{c}
\gamma\alpha\\
-(\gamma\alpha+\beta\kappa)\\
\beta\kappa
\end{array}\right];
\end{equation}
 
\begin{equation}
\frac{d}{dt}\left[\begin{array}{c}
\gamma(t)\\
\kappa(t)
\end{array}\right]=\left[\begin{array}{c}
-\beta v-(\omega_{1}-\omega_{3})\alpha\\
-\alpha v+(\omega_{2}-\omega_{3})\beta
\end{array}\right];
\end{equation}
 
\begin{equation}
\frac{d}{dt}\left[\begin{array}{c}
\alpha(t)\\
\beta(t)
\end{array}\right]=-i\Omega\left[\begin{array}{cc}
0 & -i\\
i & 0
\end{array}\right]\left[\begin{array}{c}
\alpha\\
\beta
\end{array}\right];
\end{equation}
 
\begin{equation}
\frac{d}{dt}\left[\begin{array}{c}
\Omega(t)\\
v(t)
\end{array}\right]=\left[\begin{array}{c}
0\\
\alpha\kappa+\beta\gamma
\end{array}\right];
\end{equation}

\noindent where the complex variable $\varepsilon(t)=\Omega(t)-iv(t)$.
This has a solution for the time-optimal Hamiltonian which is given
by the matrix:

\begin{equation}
\tilde{H}(t)=R\left[\begin{array}{ccc}
0 & \mathrm{cos}(\Omega t) & 0\\
\mathrm{cos}(\Omega t) & 0 & -i\mathrm{sin}(\Omega t)\\
0 & i\mathrm{sin}(\Omega t) & 0
\end{array}\right]
\end{equation}

\noindent We now apply this Hamiltonian to a wave-vector, given as
a superposition of energy eigenstates which may be (in general) time
dependent. This has the expansion:

\noindent \begin{flushleft}
\begin{equation}
\left|\Psi(t)\right\rangle =\Delta_{1}(t)\left|+R(t)\right\rangle +\Delta_{2}(t)\left|0(t)\right\rangle +\Delta_{3}(t)\left|-R(t)\right\rangle 
\end{equation}

\par\end{flushleft}

\begin{singlespace}
\noindent \begin{flushleft}
We obtain the formula for the wave-vector in the standard basis $\left|\Psi(t)\right\rangle =\left[\begin{array}{ccc}
c_{1}(t), & c_{2}(t), & c_{3}(t)\end{array}\right]^{T}$ as given by: 
\par\end{flushleft}
\end{singlespace}

\begin{flushleft}
\begin{equation}
\begin{array}{cc}
c_{1}(t)= & -\dfrac{\mathrm{sin}(\Omega't)}{\Omega'}\left[i\dfrac{z_{2}R}{\Omega'}+\Omega\left[\dfrac{z_{1}}{\Omega'}\mathrm{cos}(\Omega't)-z_{3}\mathrm{sin}(\Omega't)\right]\right]\\
 & +\mathrm{cos}(\Omega t)\left[z_{3}\mathrm{cos}(\Omega't)+z_{1}\dfrac{\mathrm{sin}(\Omega't)}{\Omega'}\right]
\end{array}
\end{equation}

\par\end{flushleft}

\begin{equation}
c_{2}(t)=\frac{1}{(\Omega')^{2}}\left[\Omega z_{2}+iR\left[z_{1}\mathrm{cos}(\Omega't)-z_{3}\mathrm{sin}(\Omega't)\right]\right]
\end{equation}

\begin{equation}
\begin{array}{cc}
c_{3}(t)= & \dfrac{\mathrm{cos}(\Omega t)}{\Omega'}\left[i\dfrac{z_{2}R}{\Omega'}+\Omega\left[z_{1}\dfrac{\mathrm{cos}(\Omega't)}{\Omega'}-z_{3}\mathrm{sin}(\Omega't)\right]\right]\\
 & -i\mathrm{sin}(\Omega t)\left[z_{3}\mathrm{cos}(\Omega't)+\dfrac{z_{1}}{\Omega'}\mathrm{sin}(\Omega't)\right]
\end{array}
\end{equation}
 where the parameters $z_{1},z_{2},z_{3}$ are given by the formulae:

\begin{equation}
\left[\begin{array}{c}
z_{1}\\
z_{2}\\
z_{3}
\end{array}\right]=\left[\begin{array}{ccc}
\Omega & -iR & 0\\
-iR & \Omega & 0\\
0 & 0 & 1
\end{array}\right]\left[\begin{array}{c}
\Delta_{2}(0)\\
\Delta_{-}(0)\\
\Delta_{+}(0)
\end{array}\right]
\end{equation}

\begin{flushleft}
\begin{equation}
\Delta_{\pm}=\frac{1}{\sqrt{2}}(\Delta_{1}\pm\Delta_{3})
\end{equation}
 
\begin{equation}
\Omega'=\sqrt{R^{2}+\Omega^{2}}
\end{equation}
 which define our initial state via: 
\par\end{flushleft}

\begin{flushleft}
\begin{equation}
\left|\Psi(0)\right\rangle =\left[\begin{array}{c}
z_{3}\\
\dfrac{1}{\Omega'^{2}}(\Omega z_{2}+iRz_{1})\\
-i\dfrac{(\Omega z_{1}+iRz_{2})}{\Omega'^{2}}
\end{array}\right];
\end{equation}
In this case we were able to find a solution despite the explicit
time dependence of the system. This example is particularly useful
as it indicates a general method of solution for the quantum brachistochrone
equation which we may use on the semisimple subgroups. 
\par\end{flushleft}

\section{Time Dependent Geodesic on SU(3)}

We now take a slightly more general Hamiltonian operator and constraint
than the previous example, being the matrices:

\begin{equation}
\tilde{H}(t)=\left[\begin{array}{ccc}
0 & \varepsilon_{1}(t) & 0\\
\varepsilon_{1}^{*}(t) & 0 & \varepsilon_{2}(t)\\
0 & \varepsilon_{2}^{*}(t) & 0
\end{array}\right];
\end{equation}

\begin{equation}
\tilde{F}(t)=\left[\begin{array}{ccc}
\omega_{1} & 0 & \kappa\\
0 & \omega_{2} & 0\\
\kappa^{*} & 0 & -(\omega_{1}+\omega_{2})
\end{array}\right]
\end{equation}

\paragraph*{\textmd{\textup{Using the quantum brachistochrone equation  we obtain
the set of coupled differential equations:}}}

\begin{equation}
\frac{d}{dt}\left[\begin{array}{c}
\kappa(t)\\
\kappa^{*}(t)
\end{array}\right]=\left[\begin{array}{c}
0\\
0
\end{array}\right]
\end{equation}

\begin{equation}
\frac{d}{dt}\left[\begin{array}{c}
\omega_{1}\\
\omega_{2}
\end{array}\right]=\left[\begin{array}{c}
0\\
0
\end{array}\right]
\end{equation}

\begin{equation}
i\frac{d}{dt}\left[\begin{array}{c}
\varepsilon_{1}(t)\\
\varepsilon_{1}^{*}(t)\\
\varepsilon_{2}(t)\\
\varepsilon_{2}^{*}(t)
\end{array}\right]=\left[\begin{array}{c}
-(\omega_{1}-\omega_{2})\varepsilon_{1}-\kappa\varepsilon_{2}^{*}\\
(\omega_{1}-\omega_{2})\varepsilon_{1}^{*}+\kappa^{*}\varepsilon_{2}\\
\kappa\varepsilon_{1}^{*}-(\omega_{1}+2\omega_{2})\varepsilon_{2}\\
-\kappa^{*}\varepsilon_{1}+(\omega_{1}+2\omega_{2})\varepsilon_{2}^{*}
\end{array}\right]
\end{equation}

\noindent from which we immediately observe that $\omega_{1},\omega_{2}$
and $\kappa$ are constants of the motion. We now assume boundary
conditions on the state, such that initially at time zero the wave-vector
is in the state $\left[\begin{array}{ccc}
1, & 0, & 0\end{array}\right]^{T}$and it is in the state $[\begin{array}{ccc}
0, & 0, & 1\end{array}]^{T}$ after a time $t=T$. We then have boundary conditions on the projection
operator as given by the matrices:

\begin{equation}
\hat{P}(0)=\left[\begin{array}{ccc}
1 & 0 & 0\\
0 & 0 & 0\\
0 & 0 & 0
\end{array}\right],\,\hat{P}(T)=\left[\begin{array}{ccc}
0 & 0 & 0\\
0 & 0 & 0\\
0 & 0 & 1
\end{array}\right]
\end{equation}

\noindent From \cite{Carlini}, we may write the boundary conditions
on the Hamiltonian as:

\begin{flushleft}
\begin{equation}
\hat{G}(t)=\left\{ \hat{G}(t),\hat{P}(t)\right\} ,0\leq t\leq T
\end{equation}
where 
\par\end{flushleft}

\begin{flushleft}
\begin{equation}
\hat{G}(t)=(\tilde{H}(t)+\tilde{F}(t))-\left\langle \psi(t)\right|(\tilde{H}(t)+\tilde{F}(t))\left|\psi(t)\right\rangle \hat{P}(t)
\end{equation}
After some elementary linear algebra, we find that the boundary conditions
are equivalent to a series of relations on the control fields: 
\par\end{flushleft}

\begin{equation}
\omega_{1}=\omega_{2}=0
\end{equation}
 
\begin{equation}
\varepsilon_{2}(0)=\varepsilon_{2}^{*}(0)=0
\end{equation}
 
\begin{equation}
\varepsilon_{1}^{*}(T)=\varepsilon_{1}(T)=0
\end{equation}
As the diagonal elements of $\tilde{F}(t)$ are constant in time,
and from the boundary conditions are initially zero, then we can conclude
that they are zero throughout the entire motion. This means that the
entire constraint matrix is constant for this situation, and we may
write the time evolution operator as: 

\begin{equation}
\hat{U}(t,0)=\mathrm{exp}(+it\tilde{F})\mathrm{\mathrm{exp}}(-it(\tilde{H}(0)+\tilde{F}))
\end{equation}

\noindent Evaluating these matrices via the eigenvalue determinant,
and Laufer's formulae\cite{Laufer} we find an explicit formula for
the state as a function of time:

\begin{equation}
\left|\psi(t)\right\rangle =\left[\begin{array}{c}
\mathrm{cos(\mathit{t\Delta})cos(\left|\kappa\right|\mathit{t\mathrm{)}+\frac{\left|\kappa\right|}{\Delta}\mathrm{sin(\left|\kappa\right|\mathit{t})}\mathrm{sin}\mathrm{(\mathit{t}\Delta)}}}\\
-\frac{i\varepsilon_{1}^{*}(0)}{\Delta}\mathrm{sin(\mathit{t}\Delta)}\\
i\kappa^{*}\left[\mathrm{\frac{1}{\left|\kappa\right|}sin(\left|\kappa\right|\mathit{t})cos(\mathit{t}\Delta)-\frac{1}{\Delta}sin(\mathit{t}\Delta)cos(\left|\kappa\right|\mathit{t})}\right]
\end{array}\right]
\end{equation}

\noindent \begin{flushleft}
where $\Delta=\sqrt{\left|\kappa\right|^{2}+\left|\varepsilon_{1}(0)\right|^{2}}$ 
\par\end{flushleft}

\noindent Calculating the projection operator, we obtain boundary
conditions on the state vector:

\begin{flushleft}
$\hat{P}(t)=\left|\psi(t)\right\rangle \left\langle \psi(t)\right|$ 
\par\end{flushleft}

\begin{flushleft}
\begin{equation}
=\left[\begin{array}{ccc}
\left|c_{1}(t)\right|^{2} & c_{1}(t)c_{2}^{*}(t) & c_{1}(t)c_{3}^{*}(t)\\
c_{1}^{*}(t)c_{2}(t) & \left|c_{2}(t)\right|^{2} & c_{2}(t)c_{3}^{*}(t)\\
c_{1}^{*}(t)c_{3}(t) & c_{2}^{*}(t)c_{3}(t) & \left|c_{3}(t)\right|^{2}
\end{array}\right]
\end{equation}

\par\end{flushleft}

\noindent After a time $t=T$ the matrix elements of the projection
operator satisfy:

\noindent 
\begin{equation}
\left|c_{1}(T)\right|^{2}=\left|c_{2}(T)\right|^{2}=0;\left|c_{3}(T)\right|^{2}=1
\end{equation}

\noindent and hence we may write the equivalent boundary conditions
as:

\begin{equation}
\left|c_{2}(T)\right|^{2}=0=(-\frac{i\varepsilon_{1}^{*}(0)}{\Delta}\mathrm{sin(\mathit{T}\Delta))^{2}}
\end{equation}

\noindent thereby obtaining the condition of time quantisation:

\begin{equation}
\mathrm{sin}(T\Delta)=0\Leftrightarrow T=\dfrac{n\pi}{\sqrt{\left|\kappa\right|^{2}+\left|\varepsilon_{1}(0)\right|^{2}}}
\end{equation}
Considering the boundary conditions on the first parameter from the
wavevector, we find immediately that $\mathrm{cos}(T\left|\kappa\right|)=0$
which implies that: 

\begin{equation}
T=\dfrac{(2n'+1)\pi}{2\left|\kappa\right|},\, n'\in\mathfrak{\mathbb{N}}
\end{equation}

\noindent For these to both be the minimum time, they must be equal.
From this we find the relationship:

\begin{equation}
\left|\kappa\right|^{2}(1-(\frac{(n'+1/2)}{n})^{2})=(\frac{(n'+1/2)}{n})^{2}\left|\varepsilon_{1}(0)\right|^{2}
\end{equation}

\noindent The quantity on the right hand side of the equality is the
square of a real number, and hence is positive. This means that the
function multiplying $\left|\kappa\right|^{2}$on the left hand side
must be positive, and hence we finally derive the minimum time for
this particular state transfer as:

\begin{equation}
\begin{array}{c}
n\geq n'+1/2\\
T_{min}=\dfrac{\pi}{2\left|\kappa\right|}=\dfrac{\sqrt{3}\pi}{2\left|\varepsilon_{1}(0)\right|}
\end{array}
\end{equation}

\noindent After some further algebra we obtain the time-optimal Hamiltonian
operator as given by the matrix:

\begin{equation}
\tilde{H}(t)=R\left[\begin{array}{ccc}
0 & \mathrm{cos}(kt) & 0\\
\mathrm{cos}(kt) & 0 & e^{-i\theta}\mathrm{sin}(kt)\\
0 & e^{+i\theta}\mathrm{sin}(kt) & 0
\end{array}\right]
\end{equation}

\noindent where $R=\left|\varepsilon_{1}(0)\right|$and $k=\left|\kappa\right|$.
Note that as the constraint was calculated to be a constant matrix,
we were able to evaluate the time evolution operator without having
to calculate the full time-dependent case. All that was required was
the time dependence of the constraint, the initial condition of the
Hamiltonian operator, and the boundary conditions on the state. This
is a very useful technique, as it allows us to move to an alternative
reference frame in which for some situations the analysis may be more
amenable to solution. Our solution is consistent with the results
of \cite{Boscain}. Much of the work that has been conducted on SU(3)
Hamiltonians has been related to STIRAP \cite{Bergmann}, \cite{Eckert},
\cite{Greentree}. We do not require their assumption of adiabaticity
nor the rotating wave approximation to derive our results. Our method
may be correctly considered to be a competing idea, as it is technically
the fastest way to achieve the desired transition as opposed to a
slow, steady process as in STIRAP.

\section{Triangular Matrices}

\noindent We now move to the more technical part of this paper, and
derive a number of results on matrix groups. The reason for this is
to develop a number of representations for angular symmetry on this
elliptic group, and also to see if there are any general formulae
which may be derived. The intent is to go from simple examples and
steadily increase in difficulty in order to solve the main problems
contained within. The space of complex matrices has a rich structure,
and the dynamics of the semigroups constitutes the behaviour of the
time dependent, periodic state. For this reason, we use a one-at-a-time
parameter approach, and show how each individual semigroup works.

\noindent The simplest way in which a set of square matrices can be
divided is into the values above the diagonal, the values below the
diagonal, and the diagonal itself. Our space of time dependent Hamiltonian
matrix operators have the standard Hermitian property, and this brings
us to the consideration of triangular matrices, both upper and lower
before we attack the main problem. Form a series of matrices $\hat{A}$:

\begin{equation}
\hat{A}(a,b,c)=\left[\begin{array}{ccc}
1 & a & b\\
0 & 1 & c\\
0 & 0 & 1
\end{array}\right]
\end{equation}
 We now input some other co-ordinate to form the matrix $\hat{A}'$:

\begin{equation}
\hat{A}'=\hat{A}(a',b',c')=\left[\begin{array}{ccc}
1 & a' & b'\\
0 & 1 & c'\\
0 & 0 & 1
\end{array}\right]
\end{equation}
The structure of the group is given by the anti-commutator and commutator,
being the fundamental symmetric and antisymmetric operators: 
\[
\left\{ \hat{A}',\hat{A}\right\} =\left[\begin{array}{ccc}
1 & 2(a+a') & (b+b')+a'c+c'a\\
0 & 1 & (b+b')+2cc'\\
0 & 0 & 1
\end{array}\right]
\]

\begin{flushleft}
\begin{equation}
=\hat{A}(a'',b'',c'')
\end{equation}

\par\end{flushleft}

\begin{flushleft}
\begin{equation}
\left[\hat{A},\hat{A}'\right]=\hat{A}\hat{A}'-\hat{A}\hat{'A}=\left[\begin{array}{ccc}
0 & 0 & 1\\
0 & 0 & 1\\
0 & 0 & 0
\end{array}\right](b-b')
\end{equation}
As we can see, these matrices form a nilpotent semigroup: 
\par\end{flushleft}

\begin{flushleft}
\begin{equation}
\left[\begin{array}{ccc}
0 & 0 & 1\\
0 & 0 & 1\\
0 & 0 & 0
\end{array}\right]\left[\begin{array}{ccc}
0 & 0 & 1\\
0 & 0 & 1\\
0 & 0 & 0
\end{array}\right]=\left[\begin{array}{ccc}
0 & 0 & 0\\
0 & 0 & 0\\
0 & 0 & 0
\end{array}\right]
\end{equation}

\par\end{flushleft}

\begin{flushleft}
\begin{equation}
\hat{A}^{2}(a,b,c)=\left[\begin{array}{ccc}
1 & a & b\\
0 & 1 & c\\
0 & 0 & 1
\end{array}\right]\left[\begin{array}{ccc}
1 & a & b\\
0 & 1 & c\\
0 & 0 & 1
\end{array}\right]
\end{equation}

\par\end{flushleft}

\begin{flushleft}
\begin{equation}
=\left[\begin{array}{ccc}
1 & 2a & 2b+ca\\
0 & 1 & 2c\\
0 & 0 & 1
\end{array}\right]=\hat{A}(2a,2b+ca,2c)
\end{equation}
Useful results: 
\begin{equation}
\hat{A}^{2}=\hat{A}+\hat{X}=\left[\begin{array}{ccc}
1 & a & b\\
0 & 1 & c\\
0 & 0 & 1
\end{array}\right]+\left[\begin{array}{ccc}
1 & a & b+ca\\
0 & 1 & c\\
0 & 0 & 1
\end{array}\right];[\hat{A},\hat{X}]=0
\end{equation}
 
\begin{equation}
\left[\begin{array}{ccc}
1 & a & b\\
0 & 1 & c\\
0 & 0 & 1
\end{array}\right]\left[\begin{array}{ccc}
1 & a & b+ca\\
0 & 1 & c\\
0 & 0 & 1
\end{array}\right]=\left[\begin{array}{ccc}
1 & 2a & 2(b+ca)\\
0 & 1 & 2c\\
0 & 0 & 1
\end{array}\right]
\end{equation}
 The exponential matrix allows us to find the time dependence of the
state, and hence solve the differential equations: 
\begin{equation}
\exp(-i\hat{A}t)=1-i\hat{A}t+\frac{1}{2!}(-i\hat{A}t)^{2}+...+\frac{1}{n!}(-i\hat{A}t)^{n}+...
\end{equation}

\par\end{flushleft}

\begin{flushleft}
\begin{equation}
\exp(-i\hat{A}t)=1-i\hat{A}t+\frac{1}{2!}(-it)^{2}(\hat{A}+\hat{X})
\end{equation}

\par\end{flushleft}

\begin{flushleft}
\begin{equation}
+\frac{1}{3!}(-it)^{3}\hat{A}(\hat{A}+\hat{X})+\frac{1}{4!}(-it)^{4}(\hat{A}+\hat{X})^{2}+...
\end{equation}
Using the determinant matrix we find the relevant polynomial eigenvalue
equation: 
\begin{equation}
\left\Vert \begin{array}{ccc}
1-\lambda & a & b\\
0 & 1-\lambda & c\\
0 & 0 & 1-\lambda
\end{array}\right\Vert =det(\hat{A}-\lambda\hat{1})=(1-\lambda)^{3}
\end{equation}
We can find a matrix which under the commutator sends upper triangular
to upper triangular: 
\begin{equation}
\left[\begin{array}{ccc}
\omega_{1} & 0 & 0\\
0 & \omega_{2} & 0\\
0 & 0 & \omega_{3}
\end{array}\right]=\hat{\Omega}
\end{equation}
 
\begin{equation}
\left[\begin{array}{ccc}
1 & a(\omega_{1}-\omega_{2}) & b(\omega_{1}-\omega_{3})\\
0 & 1 & c(\omega_{2}-\omega_{3})\\
0 & 0 & 1
\end{array}\right]=\hat{A}\hat{\Omega}-\hat{\Omega}\hat{A}
\end{equation}
We observe the determinant allows us to invert the system of equations:
\begin{equation}
det(\hat{A})=\left\Vert \begin{array}{ccc}
1 & a & b\\
0 & 1 & c\\
0 & 0 & 1
\end{array}\right\Vert =1
\end{equation}
Writing the differential equations in equivalent matrix form: 
\begin{equation}
\hat{A}\overrightarrow{x}=\left[\begin{array}{ccc}
1 & a & b\\
0 & 1 & c\\
0 & 0 & 1
\end{array}\right]\left[\begin{array}{c}
x_{1}\\
x_{2}\\
x_{3}
\end{array}\right]=\left[\begin{array}{c}
x_{1}+ax_{2}+bx_{3}\\
x_{2}+cx_{3}\\
x_{3}
\end{array}\right]
\end{equation}
 
\begin{equation}
\hat{A}\overrightarrow{x}=\frac{d\vec{x}}{dt}\leftrightarrow\left[\begin{array}{c}
x_{1}+ax_{2}+bx_{3}\\
x_{2}+cx_{3}\\
x_{3}
\end{array}\right]=\left[\begin{array}{c}
dx_{1}/dt\\
dx_{2}/dt\\
dx_{3}/dt
\end{array}\right]
\end{equation}
We solve the system: 
\par\end{flushleft}

\begin{flushleft}
\begin{equation}
\frac{d^{n}x_{1}}{dt^{n}}=x_{1}+nax_{2}+(nb+ac)x_{3}
\end{equation}
 
\begin{equation}
x_{1}(t)=\left(\frac{ac}{2}t^{2}+(bc+ak)t+\varrho\right)e^{t}
\end{equation}
 
\begin{equation}
x_{2}(t)=(k+ct)e^{t}
\end{equation}
 
\begin{equation}
x_{3}(t)=ce^{t}
\end{equation}
The dimensionality of the matrix group we are after, being the hermitian
operators can be broken down into a sum of triangular arrays. The
size of the dimension of the space is given by: 
\par\end{flushleft}

\begin{flushleft}
\begin{equation}
1+2(1)=2^{2}-1
\end{equation}
 
\begin{equation}
2+2(1+2)=3^{2}-1
\end{equation}
 
\begin{equation}
3+2(1+2+3)=4^{2}-1
\end{equation}
 
\begin{equation}
n^{2}=n+2\sum_{j=1}^{n-1}j
\end{equation}
Taking a lower triangular matrix, and evaluating the commutator with
an upper dimensional matrix, we find 
\begin{equation}
\hat{B}=\left[\begin{array}{ccc}
1 & 0 & 0\\
x & 1 & 0\\
y & z & 1
\end{array}\right]
\end{equation}
 
\begin{equation}
[\hat{A},\hat{B}]=\left[\begin{array}{ccc}
-(ax+by) & -bz & 0\\
-cy & ax-cz & bx\\
0 & ay & by+cz
\end{array}\right]
\end{equation}
Let 
\begin{eqnarray*}
x & = & a^{*},\, y=b^{*},\, z=c^{*}
\end{eqnarray*}
then we find: 
\begin{equation}
[\hat{A},\hat{B}]=\left[\begin{array}{ccc}
-(\left|a\right|^{2}+\left|b\right|^{2}) & -bc^{*} & 0\\
-cb^{*} & \left|a\right|^{2}-\left|c\right|^{2} & ba^{*}\\
0 & b^{*}a & \left|b\right|^{2}+\left|c\right|^{2}
\end{array}\right]
\end{equation}
Other than the diagonal entries, this matrix is close to the required
symmetry of the elliptic matrix operators. The relationship between
triangular matrices and Hermitian matrices is quite useful for understanding
many of the problems we deal with. Any Hermitian matrix may be given
as a sum of an upper triangular, its complex conjugate transpose,
and a diagonal matrix. We now move to consideration of the particular
Hamiltonian matrix. The Cayley-Hamilton formula for this matrix reads
as: 
\begin{equation}
\tilde{H}^{3}=(\alpha^{2}+\beta^{2})\tilde{H}
\end{equation}
and we can therefore write the propagator in the form: 
\begin{equation}
exp(-i\theta\tilde{H})=1+g_{1}(\theta)\tilde{H}+g_{2}(\theta)\tilde{H}^{2}
\end{equation}
 
\begin{equation}
\left\Vert \begin{array}{ccc}
-\lambda & \alpha & 0\\
\alpha & -\lambda & -i\beta\\
0 & i\beta & -\lambda
\end{array}\right\Vert =-\lambda^{3}+\lambda(\alpha^{2}+\beta^{2})
\end{equation}
 
\begin{equation}
\left[\tilde{H},\frac{d\tilde{H}}{dt}\right]=i(\dot{\alpha}\beta-\alpha\dot{\beta})\left[\begin{array}{ccc}
0 & 0 & 1\\
0 & 0 & 0\\
1 & 0 & 0
\end{array}\right]
\end{equation}
 
\begin{equation}
\left[\frac{d\tilde{H}}{dt},\tilde{H}^{2}\right]=(\alpha^{2}+\beta^{2})\frac{d\tilde{H}}{dt}+(\alpha\dot{\alpha}+\beta\dot{\beta})\left[\begin{array}{ccc}
0 & +\alpha & 0\\
-\alpha & 0 & i\beta\\
0 & -i\beta & 0
\end{array}\right]
\end{equation}

\par\end{flushleft}

\begin{flushleft}
\begin{equation}
\alpha\dot{\alpha}+\beta\dot{\beta}=0\Rightarrow\left[\frac{d\tilde{H}}{dt},\tilde{H}^{2}\right]=C\frac{d\tilde{H}}{dt}
\end{equation}
 
\begin{equation}
\tilde{H}(T)=\tilde{H}(0)\Rightarrow\tilde{H}(T)=\hat{U}^{\dagger}(T,0)\tilde{H}(0)\hat{U}(T,0)
\end{equation}
 
\begin{eqnarray}
\left[\tilde{H}(T),\hat{U}(T,0)\right]= & \left[\tilde{H}(0),\hat{U}(T,0)\right]= & 0
\end{eqnarray}
A periodic system has $\tilde{H}(nT)=\tilde{H}(0)$ for $n\in\mathbb{N}$,
so we obtain: 
\par\end{flushleft}

\begin{flushleft}
\begin{equation}
\left[\tilde{H}(0),\hat{U}(nT,0)\right]=0
\end{equation}
 
\begin{equation}
\hat{U}(nT,0)=e^{-in\omega T}\hat{\mathbf{1}}
\end{equation}
Initially, the time evolution operator has value unity: 
\begin{equation}
\hat{U}(0,0)=\hat{\mathbf{1}}
\end{equation}
 
\begin{equation}
i\frac{d\hat{U}}{dt}=\tilde{H}(t)\hat{U}(t,0)
\end{equation}
 
\begin{equation}
\tilde{H}(t)=R\left[\begin{array}{ccc}
0 & \mathrm{cos(}\omega t) & 0\\
\mathrm{cos(}\omega t) & 0 & -i\mathrm{sin(}\omega t)\\
0 & i\mathrm{sin(}\omega t) & 0
\end{array}\right]
\end{equation}
This operator is periodic: 
\begin{equation}
\tilde{H}[\frac{\pi}{\omega}(2n+1)]=\tilde{H}[0]
\end{equation}
with initial value: 
\begin{equation}
\tilde{H}[0]=\left[\begin{array}{ccc}
0 & R & 0\\
R & 0 & 0\\
0 & 0 & 0
\end{array}\right]
\end{equation}
Time commutating identities describe the actions of the subgroup:
\begin{equation}
\left[\tilde{H}(t),\tilde{H}(s)\right]=i(\beta(t)\alpha(s)-\beta(s)\alpha(t))\left[\begin{array}{ccc}
0 & 0 & -i\\
0 & 0 & 0\\
i & 0 & 0
\end{array}\right]
\end{equation}
We may directly integrate the Hamiltonian matrix to obtain: 
\par\end{flushleft}

\begin{flushleft}
$\int_{t}^{t+T}\tilde{H}[s]ds=$
\[
\frac{R}{\omega}\mathrm{sin}(\omega t)\left[\begin{array}{ccc}
0 & \mathrm{cos}(\omega T)-1 & 0\\
\mathrm{cos}(\omega T)-1 & 0 & -i\mathrm{sin}(\omega T)\\
0 & i\mathrm{sin}(\omega T) & 0
\end{array}\right]
\]
 
\begin{equation}
+\frac{R}{\omega}\mathrm{cos}(\omega t)\left[\begin{array}{ccc}
0 & \mathrm{sin}(\omega T) & 0\\
\mathrm{sin}(\omega T) & 0 & i(\mathrm{cos}(\omega T)-1)\\
0 & -i(\mathrm{cos}(\omega T)-1) & 0
\end{array}\right]
\end{equation}
To calculate the full time dependence of the wave-vector, we must
exponentiate the above equation, which seems daunting. However, there
are useful tricks that may be applied consistently to bring the time
evolution operator into a more useful and physically intuitive format. 
\par\end{flushleft}

\section{Invariants Of SU(3)}

The invariants of a system, in a sense, define the system itself.
Several combinations of the components of the wave-vector arise commonly
during calculations on SU(3), so by considering dynamical perspectives
from this transported reference frame we are able to see the crux
of the physics.
\begin{equation}
\tilde{H}(t)=\left[\begin{array}{ccc}
0 & \alpha & 0\\
\alpha & 0 & -i\beta\\
0 & i\beta & 0
\end{array}\right]
\end{equation}
 Redefining our system of wave-vector co-ordinates and constraints:
\begin{eqnarray*}
f_{1}=c_{1}c_{3}^{*}-c_{1}^{*}c_{3}; & f_{2}=c_{2}c_{3}^{*}-c_{2}^{*}c_{3}; & f_{3}=c_{1}c_{2}^{*}+c_{1}^{*}c_{2};
\end{eqnarray*}

\begin{flushleft}
\begin{eqnarray*}
f_{4}=c_{2}c_{3}^{*}+c_{2}^{*}c_{3}; & f_{5}=c_{1}c_{3}^{*}+c_{1}^{*}c_{3}; & f_{6}=c_{1}c_{2}^{*}-c_{1}^{*}c_{2};
\end{eqnarray*}

\par\end{flushleft}

\begin{flushleft}
\begin{equation}
\left|c_{1}\right|^{2}+\left|c_{2}\right|^{2}+\left|c_{3}\right|^{2}=1;\alpha^{2}+\beta^{2}=\mathrm{const.}
\end{equation}
 We immediately find the system of equations: 
\begin{equation}
\Re(f_{1,2})=0
\end{equation}
 
\begin{equation}
\Im(f_{3})=0
\end{equation}
 This means the domains of these functions are constrained in the
complex plane. We also find in general that: 
\begin{equation}
\frac{d}{dt}\left\langle \tilde{H}\right\rangle =\left\langle \frac{d\tilde{H}}{dt}\right\rangle 
\end{equation}
 Substituting our transformed variables into the original dynamical
equations we obtain the system: 
\begin{eqnarray}
\frac{df_{1}}{dt}=\frac{\alpha}{i}f_{4}-\beta f_{3}; &  & \frac{df_{2}}{dt}=\frac{\alpha}{i}f_{5}-2\beta\left|c_{2}\right|^{2};
\end{eqnarray}
 
\begin{eqnarray}
\frac{df_{3}}{dt}=-\beta f_{5}; &  & \frac{df_{4}}{dt}=\frac{\alpha}{i}f_{1}-2\beta\left|c_{3}\right|^{2}
\end{eqnarray}
 
\begin{eqnarray}
\frac{df_{5}}{dt}=\frac{\alpha}{i}f_{2}-\beta f_{6};
\end{eqnarray}
 
\begin{equation}
\frac{df_{6}}{dt}=2\alpha(\left|c_{2}\right|^{2}-\left|c_{1}\right|^{2})-\beta f_{2};
\end{equation}
 
\begin{equation}
\alpha\frac{df_{3}}{dt}+i\beta\frac{df_{2}}{dt}=0
\end{equation}
 Defining an observable $\hat{O}$ and our algebra of states: 
\begin{equation}
\left\langle \hat{O}\right\rangle =\left\langle \psi\left|\hat{O}\right|\psi\right\rangle 
\end{equation}
 
\begin{equation}
\left|\psi\right\rangle =\left[\begin{array}{c}
c_{1}(t)\\
c_{2}(t)\\
c_{3}(t)
\end{array}\right],\left\langle \psi\right|=\left[\begin{array}{ccc}
c_{1}^{*} & c_{2}^{*} & c_{3}^{*}\end{array}\right]=(\left|\psi\right\rangle {}^{*})^{T}
\end{equation}
 we derive the useful relations: 
\begin{equation}
\frac{d}{dt}\left\langle \hat{O}\right\rangle =\left\langle \frac{d\hat{O}}{dt}\right\rangle +i\left\langle \left[\tilde{H},\hat{O}\right]\right\rangle 
\end{equation}
 
\begin{equation}
\frac{d}{dt}\left(\hat{O}^{2}\right)=\left\{ \tilde{O},\frac{d\hat{O}}{dt}\right\} 
\end{equation}
 After some work, we obtain the fundamental dynamics: 
\[
\left\langle \left\{ \tilde{H},\frac{d\tilde{H}}{dt}\right\} \right\rangle =i(\alpha\dot{\beta}+\dot{\alpha}\beta)f_{1}
\]
 
\begin{equation}
+2(\alpha\dot{\alpha}\left|c_{1}\right|^{2}+(\alpha\dot{\alpha}+\dot{\beta}\beta)\left|c_{2}\right|^{2}+\beta\dot{\beta}\left|c_{3}\right|^{2})
\end{equation}
 
\begin{equation}
\left[\tilde{H},\tilde{H}^{2}\right]=0
\end{equation}
 
\begin{equation}
\frac{d}{dt}\left\langle \tilde{H}^{2}\right\rangle =\left\langle \left\{ \tilde{H},\frac{d\tilde{H}}{dt}\right\} \right\rangle 
\end{equation}
 The average of the energy, energy variance and energy squared may
be given in terms of these functions for this system. 
\par\end{flushleft}

\begin{flushleft}
\begin{equation}
\Delta E=\sqrt{\left\langle \dot{\psi}\left|(\mathbf{1}-\left|\psi\right\rangle \left\langle \psi\right|)\right|\dot{\psi}\right\rangle }
\end{equation}
 
\begin{equation}
\left\langle \tilde{H}\right\rangle =\alpha(c_{1}^{*}c_{2}+c_{1}c_{2}^{*})+i\beta(c_{2}c_{3}^{*}-c_{2}^{*}c_{3})
\end{equation}
 
\[
\left\langle \tilde{H}^{2}\right\rangle =\alpha^{2}(\left|c_{1}\right|^{2}+\left|c_{2}\right|^{2})+\beta^{2}(\left|c_{2}\right|^{2}+\left|c_{3}\right|^{2})
\]
 
\begin{equation}
+i\beta\alpha(c_{1}c_{3}^{*}-c_{1}^{*}c_{3)}
\end{equation}
 The total amount of probability is conserved: 
\begin{equation}
\frac{d}{dt}\left\langle \psi(t)\right.\left|\psi(t)\right\rangle =0
\end{equation}
 hence we find: 
\par\end{flushleft}

\begin{flushleft}
\begin{equation}
\beta^{2}\frac{d}{dt}\left|c_{1}\right|^{2}+\alpha^{2}\frac{d}{dt}\left|c_{3}\right|^{2}=i\beta\alpha\frac{d}{dt}(c_{1}c_{3}^{*}-c_{1}^{*}c_{3})
\end{equation}
 
\begin{equation}
\alpha\beta(\alpha c_{2}^{*}c_{3}+\beta c_{1}^{*}c_{2})=0
\end{equation}
 
\begin{eqnarray}
\alpha^{2}+\beta^{2}=R^{2} & \Rightarrow & \alpha\dot{\alpha}+\beta\dot{\beta}=0
\end{eqnarray}
 
\begin{eqnarray}
\frac{d}{dt}(\left|c_{1}\right|^{2}+\left|c_{2}\right|^{2}+\left|c_{3}\right|^{2})=0 & \Rightarrow & \frac{\alpha}{i}f_{3}+\beta f_{4}=0
\end{eqnarray}
 To calculate the probability of any particular event, we use: 
\begin{equation}
P(j|\psi)=\left|c_{j}(t)\right|^{2}=\left|\left\langle j|\psi\right\rangle \right|^{2}
\end{equation}
Obviously this is exhaustive and sums to unity.
\par\end{flushleft}

\section{Quantum Frenet-Serre Formula}

The Frenet-Serre formula of classical differential geometry relates
the path of a particle to various properties of the surface it is
rolling along, being the torsion and the curvature. This equation
has a fundamental similarity to our Hamiltonian matrix, and as such
we consider a new example, which is equivalent in all respects to
the Frenet-Serre formula. By applying our method to a problem with
known solution, we are providing a valuable check to our formalism.
Firstly, assume the dynamical system:

\begin{flushleft}
\begin{equation}
i\frac{d}{dt}\left[\begin{array}{c}
c_{1}\\
c_{2}\\
c_{3}
\end{array}\right]=\left[\begin{array}{ccc}
0 & -iK(t) & 0\\
iK(t) & 0 & -iT(t)\\
0 & iT(t) & 0
\end{array}\right]\left[\begin{array}{c}
c_{1}\\
c_{2}\\
c_{3}
\end{array}\right]
\end{equation}
Writing the explicitly parameter dependent Hamiltonian matrix and
constraint: 
\begin{equation}
\tilde{H}(t)=\left[\begin{array}{ccc}
0 & -iK(t) & 0\\
iK(t) & 0 & -iT(t)\\
0 & iT(t) & 0
\end{array}\right];\tilde{F}=\left[\begin{array}{ccc}
\omega_{1} & \alpha & \chi\\
\alpha & \omega_{2} & \beta\\
\chi^{*} & \beta & \omega_{3}
\end{array}\right]
\end{equation}
 
\begin{equation}
K,T,\alpha,\beta,\omega_{j}\in f:[0,\infty)\rightarrow\mathbb{R}
\end{equation}
 
\begin{equation}
\chi,\chi^{*}\in f:[0,\infty)\rightarrow\mathbb{C}
\end{equation}
 
\begin{equation}
\omega_{1}+\omega_{2}+\omega_{3}=0
\end{equation}
Evaluating the quantum brachistochrone equation: 
\begin{equation}
i\frac{d}{dt}\left(\tilde{H}+\tilde{F}\right)=\left[\tilde{H},\tilde{F}\right]
\end{equation}
we obtain the equivalent set of relations: 
\begin{equation}
\Im(\chi)=\mathrm{const.}=\eta
\end{equation}
 
\begin{equation}
\frac{d}{dt}(\frac{\chi-\chi^{*}}{2i})=0
\end{equation}
Hence we may write the dynamics of the curve in matrix form as: 
\begin{equation}
\frac{d}{dt}\left[\begin{array}{c}
K(t)\\
T(t)
\end{array}\right]=\left[\begin{array}{cc}
0 & -\eta\\
\eta & 0
\end{array}\right]\left[\begin{array}{c}
K(t)\\
T(t)
\end{array}\right]
\end{equation}
This admits solutions: 
\begin{equation}
T(t)=A\mathrm{sin}(\eta t)+B\mathrm{cos}(\eta t)
\end{equation}
 
\begin{equation}
K(t)=C\mathrm{sin}(\eta t)+N\mathrm{cos}(\eta t)
\end{equation}
There is no real difference between our method and the Frenet-Serre
method other than where (and if) you take the complex unit: 
\begin{equation}
\frac{d\left|\psi\right\rangle }{dt}=\hat{A}(t)\left|\psi\right\rangle 
\end{equation}
 
\begin{equation}
\tilde{H}(t)=i\hat{A}(t)
\end{equation}
The solution Hamiltonian has periodicity and initial values given
by: 
\begin{equation}
\tilde{H}(0)=\tilde{H}(\frac{2n\pi}{\eta})=i\left[\begin{array}{ccc}
0 & -N & 0\\
N & 0 & -B\\
0 & B & 0
\end{array}\right]
\end{equation}
 
\begin{equation}
\tilde{H}(\frac{n\pi}{\eta})=(-1)^{n}i\left[\begin{array}{ccc}
0 & -C & 0\\
C & 0 & -A\\
0 & A & 0
\end{array}\right]
\end{equation}
In terms of the unitary this means we can write: 
\par\end{flushleft}

\begin{flushleft}
\begin{equation}
\hat{U}(T,0)\tilde{H}(0)\hat{U}^{\dagger}(T,0)=\tilde{H}(T)=\tilde{H}(0)
\end{equation}
Our eigenvalue equation reads as: 
\begin{equation}
det(\tilde{H}-\lambda\hat{\mathbf{1}})=\left\Vert \begin{array}{ccc}
-\lambda & -iK & 0\\
iK & -\lambda & -iT\\
0 & iT & -\lambda
\end{array}\right\Vert =0
\end{equation}
Giving us the cubic polynomial: 
\begin{equation}
-\lambda^{3}+\lambda(K^{2}+T^{2})=0
\end{equation}
Now, our initial Hamiltonian was constrained to finite energy: 
\begin{equation}
Tr(\tilde{H}^{2}/2)=\mathrm{const.}
\end{equation}
Hence: 
\begin{equation}
K^{2}+T^{2}=R^{2}
\end{equation}
The solution is a circle. This fact is well known. One may consider
various different scenarios with respect to the Hamiltonian-time symmetry:
\begin{equation}
\left\langle \psi(0)\left|\tilde{H}(t)\right|\psi(0)\right\rangle =\left\langle \psi(t)\left|\tilde{H}(0)\right|\psi(t)\right\rangle 
\end{equation}
Computing this for our Hamiltonian matrix we find: 
\begin{equation}
\left\langle \psi(t)\left|\tilde{H}(0)\right|\psi(t)\right\rangle =iR\left[e^{i\omega t}z-e^{-i\omega t}z^{*}\right]
\end{equation}
 
\begin{equation}
\frac{d}{dt}\left\{ \left\langle \psi(0)\left|\tilde{H}(t)\right|\psi(0)\right\rangle \right\} =\frac{1}{i}\left\langle \psi(t)\left|\left[\tilde{H}(0),\tilde{H}(t)\right]\right|\psi(t)\right\rangle 
\end{equation}
If we rescale the time parameter, the Hamiltonian takes the simpler
form: 
\begin{equation}
\tilde{H}(t)\Rightarrow\left[\begin{array}{ccc}
0 & -i\mathrm{cos}(t) & 0\\
i\mathrm{cos}(t) & 0 & -i\mathrm{sin}(t)\\
0 & i\mathrm{sin}(t) & 0
\end{array}\right],t\Rightarrow Rt
\end{equation}
The dynamical equation of state is: 
\begin{equation}
\tilde{H}\left|\psi\right\rangle =R(C_{+}\left|+\right\rangle -C_{-}\left|-\right\rangle )
\end{equation}
which has eigenvectors given by: 
\begin{equation}
\left|+\right\rangle =\frac{1}{\sqrt{2}}\left[\begin{array}{c}
-i\mathrm{cos}(t)\\
1\\
i\mathrm{sin}(t)
\end{array}\right]
\end{equation}
 
\begin{equation}
\left|0\right\rangle =\left[\begin{array}{c}
i\mathrm{sin}(t)\\
0\\
i\mathrm{cos}(t)
\end{array}\right]
\end{equation}
 
\begin{equation}
\left|-\right\rangle =\frac{1}{\sqrt{2}}\left[\begin{array}{c}
i\mathrm{cos}(t)\\
1\\
-i\mathrm{sin}(t)
\end{array}\right]
\end{equation}
It is pleasing that this methodology delivers the standard answer
to such a well-known question. The total curvature is a constant;
the geometric figure is periodic.
\par\end{flushleft}

\section{Unitary Transformations of SU(3)}

We wish to find all the unitary transformations of SU(3), in order
to understand the geometry of states. In this case, our first step
is to examine the transformations that take the time dependent Hamiltonian
matrix from the previous worked example to a diagonal representation:

\begin{flushleft}
\begin{equation}
\tilde{H}_{FS}(t)=\hat{D}(t)\hat{L}\hat{D}^{\dagger}(t)
\end{equation}
 
\begin{equation}
\hat{L}=\left[\begin{array}{ccc}
1 & 0 & 0\\
0 & -1 & 0\\
0 & 0 & 0
\end{array}\right]
\end{equation}
 We find: 
\begin{equation}
\hat{D}(t)=\left[\begin{array}{ccc}
-\dfrac{i}{\sqrt{2}}\mathrm{cos}(t) & \dfrac{i}{\sqrt{2}}\mathrm{cos}(t) & i\mathrm{sin}(t)\\
\dfrac{1}{\sqrt{2}} & \dfrac{1}{\sqrt{2}} & 0\\
\dfrac{i}{\sqrt{2}}\mathrm{sin}(t) & -\dfrac{i}{\sqrt{2}}\mathrm{sin}(t) & i\mathrm{cos}(t)
\end{array}\right]
\end{equation}
 This has the required unitary properties: 
\begin{equation}
\hat{D}(t)\hat{D}^{\dagger}(t)=\hat{D^{\dagger}}(t)\hat{D}(t)=\hat{\mathbf{1}}
\end{equation}
 with initial value given by: 
\begin{equation}
\hat{D}(0)=\left[\begin{array}{ccc}
-\dfrac{i}{\sqrt{2}} & \dfrac{i}{\sqrt{2}} & 0\\
\dfrac{1}{\sqrt{2}} & \dfrac{1}{\sqrt{2}} & 0\\
0 & 0 & i
\end{array}\right]
\end{equation}
 We may write the time evolution of this operator in terms of an integral
equation: 
\begin{equation}
\hat{D}(t)=exp(i\hat{L}t)\hat{D}(0)-i\int_{0}^{t}\tilde{H}(s)\hat{D}(s)ds
\end{equation}
 Now, consider the eigenvectors of our elliptic Hamiltonian matrix:
\begin{equation}
\left[\begin{array}{ccc}
0 & \mathrm{cos}\phi & 0\\
\mathrm{cos}\phi & 0 & -i\mathrm{sin}\phi\\
0 & i\mathrm{sin}\phi & 0
\end{array}\right]\frac{1}{\sqrt{2}}\left[\begin{array}{c}
\mathrm{cos}\phi\\
1\\
i\mathrm{sin}\phi
\end{array}\right]=\frac{1}{\sqrt{2}}\left[\begin{array}{c}
\mathrm{cos}\phi\\
1\\
i\mathrm{sin}\phi
\end{array}\right]
\end{equation}
 
\begin{equation}
\left[\begin{array}{ccc}
0 & \mathrm{cos}\phi & 0\\
\mathrm{cos}\phi & 0 & -i\mathrm{sin}\phi\\
0 & i\mathrm{sin}\phi & 0
\end{array}\right]\left[\begin{array}{c}
-\mathrm{sin}\phi\\
0\\
i\mathrm{cos}\phi
\end{array}\right]=\left[\begin{array}{c}
0\\
0\\
0
\end{array}\right]
\end{equation}
 
\begin{equation}
\left[\begin{array}{ccc}
0 & \mathrm{cos}\phi & 0\\
\mathrm{cos}\phi & 0 & -i\mathrm{sin}\phi\\
0 & i\mathrm{sin}\phi & 0
\end{array}\right]\frac{1}{\sqrt{2}}\left[\begin{array}{c}
-\mathrm{cos}\phi\\
1\\
-i\mathrm{sin}\phi
\end{array}\right]=\frac{-1}{\sqrt{2}}\left[\begin{array}{c}
-\mathrm{cos}\phi\\
1\\
-i\mathrm{sin}\phi
\end{array}\right]
\end{equation}
 Writing a matrix with these eigenvectors as columns, we obtain: 
\begin{equation}
\hat{J}(\phi)=\left[\begin{array}{ccc}
\dfrac{1}{\sqrt{2}}\mathrm{cos}\phi & -\dfrac{1}{\sqrt{2}}\mathrm{cos}\phi & -\mathrm{sin}\phi\\
\dfrac{1}{\sqrt{2}} & \dfrac{1}{\sqrt{2}} & 0\\
\dfrac{i}{\sqrt{2}}\mathrm{sin}\phi & -\dfrac{i}{\sqrt{2}}\mathrm{sin}\phi & i\mathrm{cos}\phi
\end{array}\right]
\end{equation}
 This matrix is unitary: 
\begin{equation}
\hat{J}(\phi)\hat{J^{\dagger}}(\phi)=\hat{J^{\dagger}}(\phi)\hat{J}(\phi)=\hat{\mathbf{1}}
\end{equation}
\begin{equation}
\hat{J}(\phi)\hat{L}\hat{J^{\dagger}}(\phi)=\left[\begin{array}{ccc}
0 & \mathrm{cos}\phi & 0\\
\mathrm{cos}\phi & 0 & -i\mathrm{sin}\phi\\
0 & i\mathrm{sin}\phi & 0
\end{array}\right]
\end{equation}
 
\begin{equation}
\hat{J}(0)=\left[\begin{array}{ccc}
\dfrac{1}{\sqrt{2}} & -\dfrac{1}{\sqrt{2}} & 0\\
\dfrac{1}{\sqrt{2}} & \dfrac{1}{\sqrt{2}} & 0\\
0 & 0 & i
\end{array}\right]
\end{equation}
 and has initial value given by a Hadamard gate with a $\frac{\pi}{2}$
rotation on the ancilla. Finally, using the completely complex elliptic
Hamiltonian 
\begin{equation}
\tilde{H}=\left[\begin{array}{ccc}
0 & \epsilon_{1} & 0\\
\epsilon_{1}^{*} & 0 & \epsilon_{2}\\
0 & \epsilon_{2}^{*} & 0
\end{array}\right]
\end{equation}

\par\end{flushleft}

\begin{flushleft}
Eigenvectors: 
\begin{eqnarray}
\frac{1}{\sqrt{2}}\left[\begin{array}{c}
\mathrm{cos}\varphi\\
1\\
ie^{i\varrho}\mathrm{sin}\varphi
\end{array}\right] & \& & \left[\begin{array}{c}
ie^{-i\varrho}\mathrm{sin}\varphi\\
0\\
\mathrm{cos}\varphi
\end{array}\right]
\end{eqnarray}
 
\[
\&\frac{1}{\sqrt{2}}\left[\begin{array}{c}
-\mathrm{cos}\varphi\\
1\\
-ie^{i\varrho}\mathrm{sin}\varphi
\end{array}\right],\,\varrho=\mathrm{const.}\in[0,2\pi)
\]
 
\begin{equation}
\hat{Q}(\varphi)=\left[\begin{array}{ccc}
\dfrac{1}{\sqrt{2}}\mathrm{cos}\varphi & -\dfrac{1}{\sqrt{2}}\mathrm{cos}\varphi & ie^{-i\varrho}\mathrm{sin}\varphi\\
\dfrac{1}{\sqrt{2}} & \dfrac{1}{\sqrt{2}} & 0\\
\dfrac{i}{\sqrt{2}}e^{i\varrho}\mathrm{sin}\varphi & -\dfrac{i}{\sqrt{2}}e^{i\varrho}\mathrm{sin}\varphi & \mathrm{cos}\varphi
\end{array}\right]
\end{equation}

\par\end{flushleft}

\begin{flushleft}
\begin{equation}
\hat{Q}(0)=\left[\begin{array}{ccc}
\dfrac{1}{\sqrt{2}} & -\dfrac{1}{\sqrt{2}} & 0\\
\dfrac{1}{\sqrt{2}} & \dfrac{1}{\sqrt{2}} & 0\\
0 & 0 & 1
\end{array}\right]
\end{equation}
 The initial value of this operator is a Hadamard matrix with an ancilla
bit set to 1. It is unitary: 
\begin{equation}
\hat{Q^{\dagger}}(\varphi)\hat{Q}(\varphi)=\hat{Q}(\varphi)\hat{Q^{\dagger}}(\varphi)=1
\end{equation}
\begin{equation}
\hat{Q}(\varphi)\hat{L}\hat{Q^{\dagger}}(\varphi)=\left[\begin{array}{ccc}
0 & \mathrm{cos\varphi} & 0\\
\mathrm{cos}\varphi & 0 & -ie^{-i\varrho}\mathrm{sin}\varphi\\
0 & ie^{i\varrho}\mathrm{sin}\varphi & 0
\end{array}\right]
\end{equation}
We may summarise these formulae on unitaries into the compact identities:
\begin{equation}
S(\hat{K}[t];t)=\int_{0}^{t}\hat{K}(s)\hat{L}\hat{K}^{\dagger}(s)ds
\end{equation}
 
\begin{equation}
\dfrac{\delta S}{\delta\Lambda}=0\Rightarrow\dfrac{\delta\hat{K}}{\delta\Lambda}\hat{L}\hat{K}^{\dagger}+\hat{K}\hat{L}\dfrac{\delta\hat{K^{\dagger}}}{\delta\Lambda}=0
\end{equation}
 
\begin{equation}
\left[\tilde{H},\hat{K}\right]\hat{L}\hat{K}^{\dagger}=\hat{K}\hat{L}\left[\tilde{H},\hat{K}^{\dagger}\right]
\end{equation}
 
\begin{equation}
\left\{ \tilde{H},\hat{K}\hat{L}\hat{K}^{\dagger}\right\} =\hat{K}\left\{ \tilde{H},\hat{L}\right\} \hat{K}^{\dagger}
\end{equation}
 Our propagator for the full Hamiltonian may be written in the compact
form: 
\[
\hat{U}(\theta,0)=exp(-i\int_{0}^{\theta}ds\tilde{H}(s))
\]
\[
=\left[\begin{array}{ccc}
1+(c-1)c^{2} & -isc & -i(c-1)sc\\
-isc & c & -s^{2}\\
i(c-1)sc & s^{2} & 1+(c-1)s^{2}
\end{array}\right]
\]
 
\begin{equation}
c=\mathrm{cos}\theta,\, s=\mathrm{sin}\theta
\end{equation}
 
\begin{equation}
\hat{U}^{\dagger}(\theta,0)\hat{U}(\theta,0)=\hat{U}(\theta,0)\hat{U^{\dagger}}(\theta,0)=\hat{\mathbf{1}}
\end{equation}
 
\begin{equation}
\hat{U}(\theta,0)=\hat{\mathbf{1}}-i\mathrm{sin}\theta\tilde{H}+(\mathrm{cos}\theta-1)\tilde{H}^{2}
\end{equation}
One final useful unitary transformation is the equivalent NOT-gate,
which can be used to flip two state labels: 
\par\end{flushleft}

\begin{flushleft}
\begin{equation}
\hat{N}\left[\begin{array}{ccc}
0 & \epsilon_{1} & 0\\
\epsilon_{1}^{*} & 0 & \epsilon_{2}\\
0 & \epsilon_{2}^{*} & 0
\end{array}\right]\hat{N}^{\dagger}=\left[\begin{array}{ccc}
0 & \epsilon_{2}^{*} & 0\\
\epsilon_{2} & 0 & \epsilon_{1}^{*}\\
0 & \epsilon_{1} & 0
\end{array}\right];\hat{N}=\left[\begin{array}{ccc}
0 & 0 & 1\\
0 & 1 & 0\\
1 & 0 & 0
\end{array}\right]
\end{equation}
Writing $\left|v_{j}\right\rangle $for the j-th column of $\hat{U}$,
it is simple to show that $\left\langle v_{j}\right.\left|v_{k}\right\rangle =\delta_{jk}$.
Any rotation in this space is composed of these fundamental operators
. We now compute commutators of our fundamental transformations with
the matrix $\hat{L}$: 
\par\end{flushleft}

\begin{flushleft}
$\left[\hat{L},\hat{U}(\theta,0)\right]=$ 
\begin{equation}
\left[\begin{array}{ccc}
0 & -2i\mathrm{sin}\theta\mathrm{cos}\theta & i(\mathrm{cos}\theta-1)\mathrm{sin}\theta\mathrm{cos}\theta\\
-2i\mathrm{sin}\theta\mathrm{cos}\theta & 0 & \mathrm{sin}^{2}\theta\\
-i\mathrm{sin}\theta\mathrm{cos}\theta(\mathrm{cos}\theta-1) & -\mathrm{sin}^{2}\theta & 0
\end{array}\right]
\end{equation}
 
\begin{equation}
\left[\hat{L},\hat{J}(\phi)\right]=\left[\begin{array}{ccc}
0 & -\sqrt{2}\mathrm{cos}\phi & -\mathrm{sin}\phi\\
-\sqrt{2} & 0 & 0\\
-\frac{i}{\sqrt{2}}\mathrm{sin}\phi & -\frac{i}{\sqrt{2}}\mathrm{sin}\phi & 0
\end{array}\right]
\end{equation}
 
\begin{equation}
\left[\hat{L},\hat{Q}(\varphi)\right]=\left[\begin{array}{ccc}
0 & -\sqrt{2}\mathrm{cos}\varphi & ie^{-i\varrho}\mathrm{sin\varphi}\\
-\sqrt{2} & 0 & 0\\
-\frac{i}{\sqrt{2}}e^{i\varrho}\mathrm{sin}\varphi & -\frac{i}{\sqrt{2}}e^{i\varrho}\mathrm{sin}\varphi & 0
\end{array}\right]
\end{equation}
 
\begin{equation}
\left[\hat{L},\hat{D}(\chi)\right]=\left[\begin{array}{ccc}
0 & i\sqrt{2}\mathrm{cos}\chi & i\mathrm{sin}\chi\\
-\sqrt{2} & 0 & 0\\
-\frac{i}{\sqrt{2}}\mathrm{sin}\chi & -\frac{i}{\sqrt{2}}\mathrm{sin}\chi & 0
\end{array}\right]
\end{equation}
Hence our fundamental rotations form compact semi-groups with the
operator $\hat{L}$. One may define a sequence of polynomials via
the time evolution operator, viz.: 
\par\end{flushleft}

\begin{flushleft}
\begin{equation}
\hat{U}^{-1}(\theta,0)=\left[\dfrac{P_{l,m}(\mathrm{cos}\theta)}{q(\mathrm{cos}\theta)}\right]
\end{equation}
We have eigenreflections:
\par\end{flushleft}

\begin{flushleft}
\begin{equation}
\hat{M}_{j}=\hat{\mathbf{1}}-\left|\left.v_{k}\right\rangle \left\langle v_{k}\right.\right|
\end{equation}
 
\begin{equation}
\hat{M}_{1}=\frac{1}{2}\left[\begin{array}{ccc}
1+\mathrm{sin}^{2}\xi & -\mathrm{cos}\xi & i\mathrm{sin}\xi\mathrm{cos}\xi\\
-\mathrm{cos}\xi & 1 & i\mathrm{sin}\xi\\
-i\mathrm{sin}\xi\mathrm{cos}\xi & -i\mathrm{sin}\xi & 1+\mathrm{cos^{2}}\xi
\end{array}\right]
\end{equation}
 
\begin{equation}
\hat{M}_{2}=\left[\begin{array}{ccc}
\mathrm{sin}^{2}\xi & 0 & -i\mathrm{sin}\xi\mathrm{cos}\xi\\
0 & 1 & 0\\
i\mathrm{sin}\xi\mathrm{cos}\xi & 0 & 1+\mathrm{cos^{2}}\xi
\end{array}\right]
\end{equation}
 
\begin{equation}
\hat{M}_{3}=\frac{1}{2}\left[\begin{array}{ccc}
1+sin^{2}\xi & \mathrm{cos}\xi & i\mathrm{sin}\xi\mathrm{cos}\xi\\
\mathrm{cos}\xi & 1 & -i\mathrm{sin}\xi\\
-i\mathrm{sin}\xi\mathrm{cos}\xi & i\mathrm{sin}\xi & 1+\mathrm{cos^{2}}\xi
\end{array}\right]
\end{equation}
 These eigenflections obey the periodicity condition: 
\begin{equation}
\hat{M}_{2}(-\xi)=\hat{M}_{1}(\xi)+\hat{M}_{3}(\xi)
\end{equation}
 which is directly related to the equation of state: 
\par\end{flushleft}

\begin{flushleft}
\begin{equation}
\left|3\right\rangle _{(t+\frac{\pi}{2})}-\left|1\right\rangle _{(t+\frac{\pi}{2})}=\sqrt{2}\left|2\right\rangle _{(t)}
\end{equation}

\par\end{flushleft}

\section{Elliptic Polynomials}

\begin{flushleft}
Using the formula for the propagator, and making the substitution
$z=\mathrm{cos}\theta$ we obtain the associated z-polynomials: 
\begin{equation}
P_{l,m}(z)=q(z)\left[\hat{U}^{-1}\right]_{l,m}
\end{equation}
We display several of these functions: 
\begin{equation}
q(z)=4z^{7}-4z^{6}-8z^{5}+8z^{4}+4z^{3}-6z^{2}+1
\end{equation}
 
\begin{equation}
P_{1,1}(z)=2z^{4}-z^{3}-3z^{2}+1=p(z)
\end{equation}
 
\begin{equation}
P_{3,3}(z)=-2z^{4}+z^{3}+z^{2}-z=P(z)
\end{equation}
 
\begin{equation}
q(z)+P_{3,3}(z)=Q(z)
\end{equation}
 
\begin{equation}
Q(z)=4z^{7}-4z^{6}-8z^{5}+6z^{4}+5z^{3}-5z^{2}-z+1
\end{equation}
We may Laplace z-transform: 
\begin{equation}
L[F(z)]=L_{F}[s]=\int_{0}^{\infty}F[z(\theta)]e^{-s\theta}d\theta
\end{equation}
These transformations enable us to move between the different representations
in this space. They are analogous to the matrices used earlier to
transform to the eigenbasis. These transformed functionals are rational
polynomials with various interesting properties: 
\begin{equation}
L_{q}[s]=\frac{b_{q}(s)}{r_{q}(s)};
\end{equation}
 
\begin{eqnarray}
L_{Q}[s]=\frac{b_{Q}(s)}{r_{Q}(s)}; &  & L_{p}[s]=\frac{b_{p}(s)}{r_{p}(s)}
\end{eqnarray}
 
\begin{equation}
b_{q}(s)=s^{7}-125s^{5}+24s^{4}+192s^{3}-960s^{2}-2880s+20160
\end{equation}
 
\begin{equation}
r_{q}(s)=s^{8}
\end{equation}
 
\[
b_{Q}(s)=-2(s^{12}+120s^{10}+5016s^{8}+86527s^{6}
\]
 
\begin{equation}
+550413s^{4}+895923s^{2}+396900)
\end{equation}
 
\begin{equation}
r_{Q}=s(s^{2}+1^{2})(s^{2}+2^{2})\times\cdots\times(s^{2}+7^{2})=s\prod_{k=1}^{7}(s^{2}+k^{2})
\end{equation}

\par\end{flushleft}

\begin{flushleft}
\begin{equation}
b_{p}(s)=-(s^{8}+29s^{6}+208s^{4}+306s^{2}-144)
\end{equation}
 
\begin{equation}
r_{p}(s)=r_{P}(s)=s\prod_{k=1}^{4}(s^{2}+k^{2})
\end{equation}
 For a small circle centred on the origin: 
\begin{eqnarray}
\ointop_{\Gamma}\frac{b_{q}(z)}{r_{q}(z)}dz=1 &  & \ointop_{\Gamma}\frac{b_{p}(z)}{r_{p}(z)}dz=\frac{1}{4}
\end{eqnarray}
 
\begin{eqnarray}
\ointop_{\Gamma}\frac{b_{Q}(z)}{r_{Q}(z)}dz=-\frac{1}{2} &  & \ointop_{\Gamma}\frac{b_{P}(z)}{r_{P}(z)}dz=-\frac{1}{4}
\end{eqnarray}
 We may further Fourier transform our functions to find e.g.: 
\begin{equation}
F_{W}(k)=\int_{0}^{\infty}e^{-iks}L_{W}(s)ds
\end{equation}
 
\begin{equation}
F_{P}(k)=\int_{0}^{\infty}e^{-iks}L_{P}(s)ds
\end{equation}
 
\begin{equation}
F_{P}(k)=\dfrac{i\pi}{4}\left(u[-k]H[k]-u[k]H[-k]\right)
\end{equation}
 
\begin{equation}
u[k]=e^{4k}-e^{3k}+2e^{2k}+e^{k}+1
\end{equation}
 
\begin{equation}
H[k]=\left\{ \begin{array}{c}
0;x<0\\
1;x>0
\end{array}\right)
\end{equation}
 The other transformation we require is given by the Mellin integral:
\begin{equation}
M[\alpha]=\int_{0}^{\infty}z^{\alpha-1}f(z)dz
\end{equation}
 One immediately obtains the p-adic sequences: 
\begin{eqnarray*}
 & M_{FLP}[\alpha]=M[F_{P}(z)], & M_{FLQ}[\alpha]=M[F_{Q}(z)]
\end{eqnarray*}
 
\[
M_{FLp}[\alpha]=M[F_{p}(z)],
\]

\par\end{flushleft}

\begin{flushleft}
\begin{equation}
M_{FLP}[\alpha]=\dfrac{i\pi}{4}\Gamma(\alpha)\left\{ \frac{1}{4^{\alpha}}-\dfrac{1}{3^{\alpha}}+\dfrac{2}{2^{\alpha}}+1\right\} +\dfrac{i\pi\Xi(\alpha)}{4}
\end{equation}

\par\end{flushleft}

\begin{flushleft}
\begin{equation}
M_{FLp}[\alpha]=\dfrac{i\pi}{4}\Gamma(\alpha)\left\{ -\frac{1}{4^{\alpha}}+\dfrac{1}{3^{\alpha}}+\dfrac{2}{2^{\alpha}}+3\right\} -\dfrac{i\pi\Xi(\alpha)}{4}
\end{equation}
 
\begin{equation}
M[F_{Q}(z)]=\dfrac{i\pi}{16}\Gamma(\alpha)\left\{ -\frac{1}{7^{\alpha}}+\dfrac{2}{6^{\alpha}}+\dfrac{1}{5^{\alpha}}+\dfrac{22}{2^{\alpha}}+1\right\} +\dfrac{i\pi\Xi(\alpha)}{2}
\end{equation}
 Mellin integrating the original Laplace transform, we find for $|\alpha|<1$: 
\par\end{flushleft}

\begin{flushleft}
\begin{equation}
M_{LP}[\alpha]=\dfrac{\pi}{8}\dfrac{Z(\alpha)}{\mathrm{sin}\left(\dfrac{\pi(\alpha-1)}{2}\right)}
\end{equation}

\par\end{flushleft}

\begin{flushleft}
\begin{equation}
M_{LQ}[\alpha]=\dfrac{\pi}{96}\dfrac{Y(\alpha)}{\mathrm{sin}\left(\dfrac{\pi(\alpha-1)}{2}\right)}
\end{equation}
 
\begin{equation}
M_{Lp}[\alpha]=-\dfrac{\pi}{8}\dfrac{X(\alpha)}{\mathrm{sin}\left(\dfrac{\pi(\alpha-1)}{2}\right)}
\end{equation}
 
\begin{equation}
X(\alpha)=\left(-3+2^{\alpha}-3^{\alpha-1}+2^{2(\alpha-1)}\right)
\end{equation}
 
\begin{equation}
Y(\alpha)=(3+33\times2^{\alpha}-3^{\alpha}+3\times5^{(\alpha-1)}+6^{\alpha}-3\times7^{(\alpha-1)})
\end{equation}
 
\begin{equation}
Z(\alpha)=\left(1+2^{\alpha}-3^{\alpha-1}+2^{2(\alpha-1)}\right)
\end{equation}
The other rational function has a pole at the origin. In this description
of the space the Fourier, Laplace, Mellin integrals and cosine transform
play a complementary role to the unitary matrices used earlier. The
order of integration is compositional, in that we are fully integrating
over the variables at each time and the order must not be interchanged
except in unique circumstances. These functions, and their related
theta transforms are part of a general group of elliptic symmetry
operators. One may plot these simply on a polar plot diagram.
\par\end{flushleft}

\section{Semigroup Transformations}

\begin{flushleft}
Consider the unitary matrices $\left\{ \hat{Q},\hat{D},\hat{J}\right\} $.
We may write these matrices as the outer product of unit eigenvectors
with column vectors: 
\begin{equation}
\hat{Q}=\left[\begin{array}{ccc}
\vdots & \vdots & \vdots\\
q_{1} & q_{2} & q_{3}\\
\vdots & \vdots & \vdots
\end{array}\right]=\left[\begin{array}{ccc}
\left|q_{1}\right\rangle  & \left|q_{2}\right\rangle  & \left|q_{3}\right\rangle \end{array}\right]
\end{equation}
 
\begin{equation}
\hat{J}=\left[\begin{array}{ccc}
\vdots & \vdots & \vdots\\
j_{1} & j_{2} & j_{3}\\
\vdots & \vdots & \vdots
\end{array}\right]=\left[\begin{array}{ccc}
\left|j_{1}\right\rangle  & \left|j_{2}\right\rangle  & \left|j_{3}\right\rangle \end{array}\right]
\end{equation}

\par\end{flushleft}

\begin{flushleft}
\begin{equation}
\hat{D}=\left[\begin{array}{ccc}
\vdots & \vdots & \vdots\\
d_{1} & d_{2} & d_{3}\\
\vdots & \vdots & \vdots
\end{array}\right]=\left[\begin{array}{ccc}
\left|d_{1}\right\rangle  & \left|d_{2}\right\rangle  & \left|d_{3}\right\rangle \end{array}\right]
\end{equation}
Writing $\left|A_{l}(s)\right\rangle _{m}=\left[\hat{A}(s)\right]_{l,m}=\Psi_{l,m}^{A}(s)$.
We first consider the various transformations on the unitaries: 
\begin{equation}
\hat{L}^{a_{i}}(\alpha)\left|a_{i}(\sigma)\right\rangle =\left|a_{i}(\sigma+\alpha)\right\rangle 
\end{equation}
Finding the matrix operators: 
\[
\left\{ \hat{L}^{D}\right\} =\left[\begin{array}{ccc}
\mathrm{cos}\sigma & 0 & \mathrm{sin}\sigma\\
0 & 1 & 0\\
-\mathrm{sin}\sigma & 0 & \mathrm{cos}\sigma
\end{array}\right]\&\left[\begin{array}{ccc}
\mathrm{cos}\sigma & 0 & -\mathrm{sin}\sigma\\
0 & 1 & 0\\
+\mathrm{sin}\sigma & 0 & \mathrm{cos}\sigma
\end{array}\right]
\]
 
\begin{equation}
\end{equation}
 
\begin{eqnarray*}
\left\{ \hat{L}^{Q}\right\} =\left[\begin{array}{ccc}
\mathrm{cos}\sigma & 0 & ie^{-i\theta}\mathrm{sin}\sigma\\
0 & 1 & 0\\
ie^{i\theta}\mathrm{sin}\sigma & 0 & \mathrm{cos}\sigma
\end{array}\right],
\end{eqnarray*}

\par\end{flushleft}

\begin{flushleft}
\begin{equation}
\&\left[\begin{array}{ccc}
\mathrm{cos}\sigma & 0 & ie^{-i\theta}\mathrm{sin}\sigma\\
0 & 1 & 0\\
-ie^{i\theta}\mathrm{sin}\sigma & 0 & \mathrm{cos}\sigma
\end{array}\right]
\end{equation}
 
\[
\left\{ \hat{L}^{J}\right\} =\left[\begin{array}{ccc}
\mathrm{cos}\sigma & 0 & i\mathrm{sin}\sigma\\
0 & 1 & 0\\
i\mathrm{sin}\sigma & 0 & \mathrm{cos}\sigma
\end{array}\right]\&\left[\begin{array}{ccc}
\mathrm{cos}\sigma & 0 & -i\mathrm{sin}\sigma\\
0 & 1 & 0\\
-i\mathrm{sin}\sigma & 0 & \mathrm{cos}\sigma
\end{array}\right]
\]
 
\begin{equation}
\&\left[\begin{array}{ccc}
\mathrm{cos}\sigma & 0 & -i\mathrm{sin}\sigma\\
0 & 1 & 0\\
i\mathrm{sin}\sigma & 0 & \mathrm{cos}\sigma
\end{array}\right]
\end{equation}

\par\end{flushleft}

\section{Fermat Principle}

Take the action function given as the classical Fermat principle:

\begin{equation}
S(y_{\alpha},\dot{y}_{\alpha}|t')=\int_{0}^{t'}n(y_{\alpha})\sqrt{\dot{y}_{\alpha}^{2}}dt
\end{equation}
The Euler-Lagrange equations then read as:
\begin{equation}
\sqrt{\dot{y}_{\alpha}^{2}}\dfrac{\partial n}{\partial y_{\alpha}}-\dfrac{d}{ds}\left(\dfrac{\dot{y}_{\alpha}n(y_{\alpha})}{\sqrt{\dot{y}_{\alpha}^{2}}}\right)=0
\end{equation}
We consider the particular case of one co-ordinate: 
\begin{equation}
n(y_{\alpha})=n(y)
\end{equation}
and transform the co-ordinate as $y=ia\sin\phi$. The differential
equation then reads as: 
\begin{equation}
i\sqrt{a^{2}+y^{2}}\dfrac{\partial n}{\partial y}\dfrac{\partial\phi}{\partial t}-\dfrac{dn(y)}{ds}=0
\end{equation}
We may therefore write, after taking the appropriate limit, that:
\begin{equation}
\dfrac{\partial n}{\partial\phi}(i\sqrt{a^{2}+y^{2}}\dfrac{\partial\phi}{\partial t}\dfrac{\partial y}{\partial\phi}-1)=0
\end{equation}
In general, $\dfrac{\partial n}{\partial\chi}\neq0$, so we therefore
have the parametric identity: 
\begin{equation}
\dfrac{\partial\phi}{\partial t}=\dfrac{-i}{\sqrt{a^{2}+y^{2}}}\dfrac{\partial\phi}{\partial y}
\end{equation}
Applying this twice, we find the elliptic differential equation:
\begin{equation}
\dfrac{\partial^{2}\phi}{\partial t^{2}}=-\dfrac{1}{\sqrt{a^{2}+y^{2}}}\dfrac{\partial}{\partial y}(\dfrac{1}{\sqrt{a^{2}+y^{2}}}\dfrac{\partial\phi}{\partial y})
\end{equation}
\begin{equation}
(a^{2}+y^{2})\dfrac{\partial^{2}\phi}{\partial t^{2}}=-\dfrac{\partial^{2}\phi}{\partial y^{2}}+\dfrac{y}{\sqrt{a^{2}+y^{2}}}\dfrac{\partial\phi}{\partial y}
\end{equation}

\section{Elliptic Wave Equations}

\begin{flushleft}
Writing $\left|A_{l}(s)\right\rangle _{m}=\left[\hat{A}(s)\right]_{l,m}=\Psi_{l,m}^{A}(s)$,
we now consider various operators on the respective wave-functions.
It is important to note that the variable may properly be held to
be a periodic complex variable, and the function extended meromorphically
to the complex space. We will work only on the section where the wave
is continuous, invertible and free of poles, so we may multiply and
divide by trigonometric functions. In this section, as we are dealing
with changes of variables, we must use partial differential operators
to distinguish them from the strict time derivative treated elsewhere
in the paper. Let us examine a particular co-ordinate transform, to
show how we may write down the wave equations. We have:
\par\end{flushleft}

\begin{equation}
\left[D(\chi)\right]_{3,1}=\dfrac{i}{\sqrt{2}}\mathrm{sin}(\chi)=\tau_{3,1}^{D}=\tau
\end{equation}
where we drop the subscripts for convenience. Directly differentiating:

\[
\dfrac{d\tau}{d\chi}=\dfrac{i}{\sqrt{2}}\mathrm{cos}\chi
\]
Squaring the derivative:

\[
(d\tau)^{2}=-(\frac{1}{2}+\tau^{2})d\chi^{2}
\]
We find after inverting the fraction and taking the square root:

\begin{equation}
\dfrac{d\chi}{d\tau}=\dfrac{\partial\chi}{\partial\tau}=\dfrac{-i}{\sqrt{\dfrac{1}{2}+\tau^{2}}}
\end{equation}
Partial differentiating a function twice with respect to our parameter:

\begin{equation}
\dfrac{\partial^{2}\Psi}{\partial\tau^{2}}=\left(\dfrac{\partial\chi}{\partial\tau}\right)^{2}\dfrac{\partial^{2}\Psi}{\partial\chi^{2}}+\dfrac{\partial\Psi}{\partial\chi}\dfrac{\partial}{\partial\tau}\left(\dfrac{\partial\chi}{\partial\tau}\right)
\end{equation}
Substitution of the appropriate values brings:

\begin{equation}
\dfrac{\partial^{2}\Psi}{\partial\tau^{2}}=\dfrac{1}{\frac{1}{2}+\tau^{2}}\left(-\dfrac{\partial^{2}\Psi}{\partial\chi^{2}}+\dfrac{i\tau}{\sqrt{\frac{1}{2}+\tau^{2}}}\dfrac{\partial\Psi}{\partial\chi}\right)
\end{equation}

\noindent \begin{flushleft}
We can then find the differential-integral equations for the wave-function.
In general it takes the form: 
\par\end{flushleft}

\begin{flushleft}
\begin{equation}
\Psi_{i,j}^{k}[\sigma]=\Psi_{i,j}^{k}[0]+\epsilon_{i,j}^{k}\int_{0}^{\sigma}d\nu.\hat{V}(\nu)\Psi_{i,j}^{k}[\nu]
\end{equation}
 
\begin{equation}
\hat{V}(\nu)=F_{1}^{i,j,k}(\nu)\left(\mu_{i,j,k}F_{2}^{i,j,k}(\nu)+\beta_{i,j}^{k}\nabla_{\nu}^{2}\right)
\end{equation}
 
\begin{equation}
\epsilon_{i,j}^{k}\in\{1,-1,i,-i\};\beta_{i,j}^{k}\in\{1,-1\};\nabla_{u}^{2}=\dfrac{\partial^{2}}{\partial u^{2}}
\end{equation}
 
\begin{equation}
(F_{1}^{i,j,k},F_{2}^{i,j,k})\in\left\{ (\mathrm{cot}\nu,\mathrm{cos^{2}}\nu),(\mathrm{tan}\nu,\mathrm{sin^{2}}\nu)\right\} 
\end{equation}
 Our entire wavefunction is then: 
\begin{equation}
\Psi_{i,j}^{k}[s,\tau_{i,j}^{k}]=\Gamma_{i,j}^{k}[s]\Psi_{i,j}^{k}[\tau_{i,j}^{k}]
\end{equation}
 
\begin{equation}
\dfrac{\partial^{2}\Gamma_{i,j}^{k}[s]}{\partial s^{2}}=\mu_{i,j,k}\Gamma_{i,j}^{k}[s]
\end{equation}
These formulae are generated by a group of partial differential equations
of type: 
\begin{equation}
(z^{2}+a^{2})\dfrac{\partial^{2}\Psi}{\partial z^{2}}=-\dfrac{\partial^{2}\Psi}{\partial\chi^{2}}+\dfrac{iz}{\sqrt{z^{2}+a^{2}}}\dfrac{\partial\Psi}{\partial\chi}
\end{equation}
 
\begin{equation}
z=f(\chi),|z|<a,f(\chi+m\omega)=f(\chi)
\end{equation}
and we take the angular variable so as to avoid any discontinuities
in the relevant trigonometric ratios other than at the extremum, where
we are guaranteed analytic continuity by the hyperbolic sine-sine
relationship. They are on the perimeter of the wavefunction; we are
guaranteed continuity and unitarity by virtue of our matrix construction.
One can label the functions by their $F_{1}^{i,j,k},\epsilon_{i,j}^{k}$
and $\beta_{i,j}^{k}$ . If we consider these functions as proper
analytic functions in the complex plane, it is straightforward to
derive various differential systems, analogous to the Cauchy-Riemann
equations. For example, if $z=re^{i\theta}$, $f(z)=u(z)+iv(z)$,
we obtain the cylindrical polar set:
\begin{equation}
\dfrac{\partial u}{\partial r}=\dfrac{1}{r}\dfrac{\partial v}{\partial\theta}
\end{equation}
\begin{equation}
\dfrac{\partial v}{\partial r}=-\dfrac{1}{r}\dfrac{\partial u}{\partial\theta}
\end{equation}
\begin{equation}
\dfrac{\partial^{2}f}{\partial r^{2}}+\dfrac{1}{r}\dfrac{\partial f}{\partial r}+\dfrac{1}{r^{2}}\dfrac{\partial^{2}f}{\partial\theta^{2}}=0
\end{equation}
If $z=r\cos\theta$, $f(z)=u(z)+iv(z)$, we find the symmetrical equations:
\begin{equation}
\dfrac{\partial u}{\partial r}=-\cot\theta\dfrac{\partial v}{\partial\theta}
\end{equation}
\begin{equation}
\dfrac{\partial v}{\partial r}=-\cot\theta\dfrac{\partial u}{\partial\theta}
\end{equation}
If $z=ir\cos\theta$, $f(z)=u(z)+iv(z)$, we find polar elliptic equations:
\begin{equation}
r\dfrac{\partial u}{\partial r}=-\dfrac{z}{\sqrt{z^{2}+r^{2}}}\dfrac{\partial v}{\partial\theta}
\end{equation}
\begin{equation}
r\dfrac{\partial v}{\partial r}=\dfrac{z}{\sqrt{z^{2}+r^{2}}}\dfrac{\partial u}{\partial\theta}
\end{equation}
Differentiating the correct parameterisation, analytically continuing
it to the complex plane and using separation of variables allows us
to solve for the ground state of the system, which can be quite difficult.
The radial part of the differential equation is generally the circular
Bessel function:
\par\end{flushleft}

\begin{flushleft}
\begin{equation}
\dfrac{d^{2}R(r)}{dr^{2}}+\dfrac{1}{r}\dfrac{dR(r)}{dr}+\dfrac{\alpha}{r^{2}}R(r)=0
\end{equation}
The angular part of the particular solution is then:
\begin{equation}
\Psi(r,\theta)=R(r)Y(\theta)
\end{equation}
\begin{equation}
\dfrac{d^{2}Y(\theta)}{d\theta^{2}}+F_{1}(\theta)\dfrac{dY(\theta)}{d\theta}+\alpha F_{2}(\theta)Y(\theta)=0
\end{equation}

\par\end{flushleft}

\section{Integral Relations }

Expanding the Hamiltonian matrix in the spectral representation, we
find:

\begin{equation}
\tilde{H}=\sum_{n}E_{n}\left|\phi_{n}\right\rangle \left\langle \phi_{n}\right|
\end{equation}
In this particular situation, the eigenenergies are equally separated
from zero, we may therefore write:
\begin{equation}
\tilde{H}=E(\left|\phi_{+}\right\rangle \left\langle \phi_{+}\right|-\left|\phi_{-}\right\rangle \left\langle \phi_{-}\right|)
\end{equation}
An alternative way of writing this Hamiltonian is:
\begin{equation}
\tilde{H}=\sum_{i,j}\epsilon_{ij}(t)\hat{Q}_{ij}+\epsilon_{ij}^{*}(t)\hat{Q}_{ij}^{T}
\end{equation}
\begin{equation}
\hat{Q}_{ij}=\left|i\right\rangle \left\langle j\right|
\end{equation}
Using the dynamical equations of motion, 
\begin{equation}
i\dfrac{d}{dt}\left[\begin{array}{c}
\left|\phi_{+}\right\rangle \\
\left|\phi_{-}\right\rangle 
\end{array}\right]=E\left[\begin{array}{cc}
1 & 0\\
0 & -1
\end{array}\right]\left[\begin{array}{c}
\left|\phi_{+}\right\rangle \\
\left|\phi_{-}\right\rangle 
\end{array}\right]
\end{equation}
Let us now expand the wave-vector in the eigenbasis:
\begin{equation}
i\dfrac{d}{dt}(\sum_{j}c_{j}(t)\left|j(t)\right\rangle )=\sum_{j}E_{j}\left|j(t)\right\rangle 
\end{equation}
Dropping the sums for convenience and assuming Einstein summation
convention, we evaluate the differential equation:
\begin{equation}
ic_{j}(t)\dfrac{d}{dt}\left|j(t)\right\rangle =(E_{j}-i\dfrac{dc_{j}}{dt})\left|j(t)\right\rangle 
\end{equation}
Consider a path integral on our state space:
\begin{equation}
\left[K_{jk}(T,0)\right]=\sqiintop_{\check{x}_{j}(0)}^{\check{x}_{k}(T)}exp(-i\int_{0}^{T}ds\hat{\mathcal{L}}[E_{j},c_{j}|s])D[\check{x}_{\alpha}(s)]
\end{equation}
We may write the Lagrangian density as the formula:
\begin{equation}
\hat{\mathcal{L}}=-E_{j}\dot{c}_{j}(s)\delta_{jk}+ic_{j}(s)\left\langle k(s)\right|\widetilde{H}(s)\left|j(s)\right\rangle 
\end{equation}
Up to isomorphism, this fulfils an equivalent role to the time evolution
operator. As we have managed to already find a number of differential
relationships, the necessary question is to enquire as to whether
there are a series of equivalent integral identities that we may use
to evaluate these particular groups of elliptic differential equations.
Consider a simple toy problem of a wave-packet interacting with a
time dependent oscillating potential:
\begin{equation}
i\hbar\dfrac{\partial\Psi(x,t)}{\partial t}=-\dfrac{\hbar^{2}}{2m}\dfrac{\partial^{2}\Psi(x,t)}{\partial x^{2}}+V_{0}\cos\omega t.\Psi(x,t)
\end{equation}
Using natural units and rescaling the time, we may write this as a
neat operator equation:
\begin{equation}
i\dfrac{\partial\Psi(x,t)}{\partial t}=(\hat{H}_{0}+\hat{H}_{I}(t))\Psi(x,t)
\end{equation}
Now, if we cosine-transform the time co-ordinate, our equations of
motion transform:

\begin{equation}
z=\cos\omega t
\end{equation}
\begin{equation}
t=\dfrac{1}{\omega}\cos^{-1}(z)
\end{equation}
\begin{equation}
\hat{H}_{I}(t)\mapsto\hat{H}_{I}(z)=V_{0}z
\end{equation}
\begin{equation}
\hat{H}_{0}\mapsto\hat{H}_{0}
\end{equation}
\begin{equation}
\dfrac{\partial}{\partial t}\mapsto\dfrac{\partial z}{\partial t}\dfrac{\partial}{\partial z}=\omega\sqrt{1-z^{2}}\dfrac{\partial}{\partial z}
\end{equation}
Writing our original wave-equation in the transformed frame, we find:
\begin{equation}
i\omega\sqrt{1-z^{2}}\dfrac{\partial\Psi(x,z)}{\partial z}=-\dfrac{1}{2}\dfrac{\partial^{2}\Psi(x,z)}{\partial x^{2}}+V_{0}z\Psi(x,z)
\end{equation}
Using the method of separation of variables, we write an ansatz for
our wavefunction:
\begin{equation}
\Psi(x,z)=\phi(x)\beta(z)
\end{equation}
Finding the equivalent set of equations:
\begin{equation}
E\phi(x)=-\dfrac{1}{2}\dfrac{d^{2}\phi(x)}{dx^{2}}
\end{equation}
\begin{equation}
i\omega\dfrac{d\beta(z)}{dz}=\dfrac{(E-V_{0}z)}{\sqrt{1-z^{2}}}\beta(z)
\end{equation}
The first of these equations is readily solvable using complex exponentials,
however, the second presents us with some difficulties. Firstly, we
rescale the variables to place the differential equation into dimensionless
form, finding:
\begin{equation}
i\dfrac{d\beta(z)}{dz}=\dfrac{(1-\alpha z)}{\sqrt{1-z^{2}}}\beta(z)
\end{equation}
Directly differentiating:
\begin{equation}
(1-z^{2})\dfrac{d^{2}\beta}{dz^{2}}=(p(\alpha,z)-\dfrac{iq(\alpha,z)}{\sqrt{1-z^{2}}})\beta
\end{equation}
where the p's and q's are polynomial functions. This is one of our
basic elliptic differential equations. Note the useful integral formula:

\begin{equation}
\intop_{a}^{b}\sec z(1+\tan z)dz=\intop_{y_{a}}^{y_{b}}\dfrac{dy}{\sqrt{y^{2}-1}}+i(y_{b}-y_{a})
\end{equation}
where $y=\sec z$. To consider further transformations, consider first
one of the polynomials from the previous section: 
\begin{equation}
b_{1}(z)=z^{8}+29z^{6}+208z^{4}+306z^{2}-144
\end{equation}
It is possible to completely specify this polynomial by its root set,
viz.:
\begin{equation}
b_{1}(z)=0,z\in\{\pm\alpha_{1},\pm i\beta_{1},\pm i\delta_{1},\pm i\gamma_{1}\}
\end{equation}
Moving sequentially through the other relevant polynomials, we find:
\begin{equation}
b_{2}(z)=z^{7}-125z^{5}+25z^{4}+192z^{3}-960z^{2}-2880z+20160
\end{equation}
\begin{equation}
b_{2}(z)=0,z\in\{\alpha_{2},\beta_{2},-\gamma_{2},z_{1},z_{1}^{*},z_{2},z_{2}^{*}\}
\end{equation}
This point set has the neat property of being a pentagram in the complex
plane with three non-symmetrical points on the real axis, one of which
is the primary vertice of the pentagon.
\[
b_{3}(z)=z^{6}+120z^{5}+5016z^{4}+86527z^{3}
\]
\begin{equation}
+550413z^{2}+896923z+396900
\end{equation}
This polynomial has its root set confined to the negative real axis:
\begin{equation}
b_{3}(z)=0,z\in\{-\alpha_{3},-\beta_{3},-\gamma_{3},-\delta_{3},-\epsilon_{3},-\zeta_{3}\}
\end{equation}
Finally, consider the polynomial $b_{4}(z)=b_{1}(\sqrt{z}).$As the
original polynomial is even in z, the new square-root transformed
polynomial is of the form:
\begin{equation}
b_{4}(z)=z^{4}+29z^{3}+208z^{2}+306z-144
\end{equation}
 which has a root set:
\begin{equation}
b_{4}(z)=0,z\in\{\alpha_{4},-\beta_{4},-\gamma_{4},-\delta_{4}\}
\end{equation}
By examining the cosine and exponentially transformed polynomials,
$b_{j}(\cos\theta)$ \& $b_{j}(\exp i\theta)$, and real and imaginary
parts thereof, it is a simple exercise to develop a series of polar
representations for these polynomial functions. The complex analytic
behaviour is generated by these or similar sets of roulettes. They
are related to a number of interesting integral relations. We begin
with a simple example: 
\begin{equation}
\int_{0}^{2\pi}\dfrac{b_{4}(u)du}{\sqrt{1-u^{2}}}=ia+Q(\pi)\sqrt{4\pi^{2}-1}
\end{equation}
The other polynomials have similar identities with respect to this
measure:
\begin{equation}
\int_{-1}^{+1}\dfrac{b_{1}(u)du}{\sqrt{1-u^{2}}}=\dfrac{12331\pi}{128}
\end{equation}
\begin{equation}
\int_{0}^{3\pi}b_{1}(u)\sqrt{1-u^{2}}du=b\pi+i\sqrt{9\pi^{2}-1}(c+(9\pi^{2}-1)P(\pi))
\end{equation}
\begin{equation}
P(\pi)=\sum_{n=0}^{3}C_{n}\pi^{2n+1}
\end{equation}
\begin{equation}
\int_{0}^{2\pi}b_{1}(u)\sqrt{1-u^{2}}du=k\pi+i\sqrt{4\pi^{2}-1}(l+(4\pi^{2}-1)Y(\pi))
\end{equation}
\begin{equation}
Y(\pi)=\sum_{n=0}^{3}D_{n}\pi^{2n+1}
\end{equation}
\begin{equation}
\{D_{n}\}\subseteq\{C_{n}\},D_{n}=0\mathrm{\, mod}C_{j}
\end{equation}
\begin{equation}
\int_{0}^{\pi/2}b_{1}(u)\sqrt{1-u^{2}}du=g\pi+i\sqrt{\pi^{2}-4}(f+(\pi^{2}-4)r(\pi))
\end{equation}
\begin{equation}
r(\pi)=\sum_{n=0}^{3}M_{n}\pi^{2n+1}
\end{equation}

\begin{equation}
\{M_{n}\}\subseteq\{C_{n}\}
\end{equation}
\begin{equation}
\int_{-l}^{+l}b_{1}(u)\sqrt{1-u^{2}}du=\sqrt{1-l^{2}}(\sum_{0}^{3}C_{n}l^{2n+1})-m\sin^{-1}(l)
\end{equation}
\begin{equation}
\int_{0}^{a}\dfrac{b_{1}(u)du}{\sqrt{1-u^{2}}}=ip(a)\sqrt{a^{2}-1}+\lambda a+i\sin^{-1}(a)
\end{equation}
Off the unit circle, these integrals lose analyticity unless we modify
the weighting function. One way to do this is by defining a co-distribution:
\begin{equation}
\int_{-1}^{+1}du(1-u^{2})^{\alpha/2}=\sqrt{\pi}\dfrac{\Gamma(\dfrac{\alpha}{2}+1)}{\Gamma(\dfrac{\alpha}{2}+\dfrac{3}{2})}
\end{equation}
We then form marginal distributions:
\begin{equation}
p_{k,W}(\alpha)=\dfrac{\int_{\Omega}p_{k}(s)W(s,\alpha)ds}{\int_{\Omega}W(s,\alpha)ds}
\end{equation}
\begin{equation}
b_{1,W}(\alpha)=-3\dfrac{\sum q_{n}\alpha^{n}}{\prod(\alpha+2(n+1)+1)}=-3\dfrac{p(\alpha)}{q(\alpha)}
\end{equation}
Using partial fractions to expand the denominator:
\[
\dfrac{1}{\prod(\alpha+2(n+1)+1)}=\dfrac{1}{48}(\dfrac{1}{\alpha+3}-\dfrac{1}{\alpha+9})
\]
\begin{equation}
+\dfrac{1}{16}(\dfrac{1}{\alpha+7}-\dfrac{1}{\alpha+5})
\end{equation}
\begin{equation}
p(\alpha)=48\alpha^{4}+1050\alpha^{3}+7538\alpha^{2}+17653\alpha-1214
\end{equation}
\begin{equation}
\lim_{\alpha\rightarrow\infty}b_{1,W}(\alpha)=-144
\end{equation}
This is not the only co-distribution, we may form an equivalent functional:
\begin{equation}
W'(\alpha,s)=(1-\alpha^{2})^{-s/2}=W(-s,\alpha)
\end{equation}
These integral relationships are then neatly expressible in terms
of hypergeometric functions:
\begin{equation}
\int d\alpha W(-s,\alpha)=\alpha\left\{ \begin{array}{cc}
\dfrac{1}{2} & \alpha^{2}\\
\dfrac{s}{2} & \dfrac{3}{2}
\end{array}\right\} 
\end{equation}
\begin{equation}
\int d\alpha W(-s,\alpha)b_{1,W}(\alpha)=\sum_{n=1}^{5}\left\{ \begin{array}{cc}
1/2 & \alpha^{2}\\
\dfrac{s}{2} & \dfrac{n}{2}+1
\end{array}\right\} \alpha^{n}C_{n}
\end{equation}
The point symmetry group of $p(\alpha)$ is a kite, all the zeroes
of $q(\alpha)$ lie along the negative real axis.

\section{Chebyshev Polynomials}

Consider a unit triangle in the complex plane. Using De-Moivre's theorem:

\begin{equation}
(e^{i\theta})^{m}=(\cos\theta+i\sin\theta)^{m}=\cos m\theta+i\sin m\theta
\end{equation}
We may write an alternative transformation, viz.:
\begin{equation}
z=x+iy=x+i\sqrt{1-x^{2}}
\end{equation}
The left-hand side of De-Moivre's theorem then reads as:
\begin{equation}
(x+i\sqrt{1-x^{2}})^{m}=\cos m\theta+i\sin m\theta
\end{equation}
We define the Chebyshev polynomial of the first kind:
\begin{equation}
T_{m}(\cos\theta)=\cos m\theta
\end{equation}
Using the double angle formula, it is a simple exercise to derive
the first recursion relation:
\begin{equation}
T_{m}(x)=xT_{m-1}(x)+(x^{2}-1)U_{m-1}(x)
\end{equation}
where we have defined the Chebyshev polynomial of the second kind:
\begin{equation}
U_{m}(\cos\theta)=\dfrac{\sin m\theta}{\sin\theta}
\end{equation}
with recursion relation:
\begin{equation}
U_{m}(x)=xU_{m-1}(x)+T_{m-1}
\end{equation}
Directly differentiating the recursion relations, we obtain the system:
\begin{equation}
\dfrac{dT_{m}(x)}{dx}=mU_{m}(x)
\end{equation}
\begin{equation}
(1-x^{2})\dfrac{dU_{m}}{dx}=m(xU_{m}-mT_{m})
\end{equation}
\begin{equation}
(1-x^{2})\dfrac{d^{2}T_{m}}{dx^{2}}=(mxU_{m}-m^{2}T_{m})
\end{equation}
\begin{equation}
(1-x^{2})\dfrac{d^{2}T_{m}}{dx^{2}}-x\dfrac{dT_{m}}{dx}+m^{2}T_{m}=0
\end{equation}
This differential equation is related to the elliptic wave equations
used in the previous sections. This comes about due to our particular
break-down of the wave-function in terms of roots of unity on the
unit circle.

\section{Circular Co-ordinates}

Consider the standard cylindrical co-ordinate system in two dimensions,
i.e. a circular system:
\begin{equation}
\vec{x}=r\left(\begin{array}{c}
\cos\theta\\
\sin\theta
\end{array}\right)
\end{equation}
Directly differentiating to find the velocity vector:
\begin{equation}
\dfrac{d\vec{x}}{dt}=\dot{r}\hat{e}_{r}+r\dot{\theta}\hat{e}_{\theta}
\end{equation}
The polar unit vectors have representation:
\begin{equation}
\hat{e}_{r}(\theta)=\left(\begin{array}{c}
\cos\theta\\
\sin\theta
\end{array}\right),\,\hat{e}_{\theta}(\theta)=\left(\begin{array}{c}
-\sin\theta\\
\cos\theta
\end{array}\right)
\end{equation}
\begin{equation}
\hat{e}_{r}(\theta)\cdot\hat{e}_{\theta}(\theta)=0
\end{equation}
\begin{equation}
\hat{e}_{r,\theta}(\theta+\alpha)=\hat{L}(\alpha)\hat{e}_{r,\theta}(\theta)
\end{equation}
\begin{equation}
\hat{L}(\alpha)=\left[\begin{array}{cc}
\cos\alpha & -\sin\alpha\\
\sin\alpha & \cos\alpha
\end{array}\right]
\end{equation}
Writing out the arc-length for the system:
\begin{equation}
ds^{2}=dr^{2}+r^{2}d\theta^{2}
\end{equation}
we derive the gradient for the wave-function:
\begin{equation}
\nabla\Psi=\hat{e}_{r}\dfrac{\partial\Psi}{\partial r}+\dfrac{\hat{e}_{\theta}}{r}\dfrac{\partial\Psi}{\partial\theta}
\end{equation}
Deriving the Laplacian in circular co-ordinates:
\begin{equation}
\nabla^{2}\Psi=(\nabla\cdot\nabla)\Psi=\dfrac{\partial^{2}\Psi}{\partial r^{2}}+\dfrac{1}{r}\dfrac{\partial\Psi}{\partial r}+\dfrac{1}{r^{2}}\dfrac{\partial^{2}\Psi}{\partial\theta^{2}}
\end{equation}
The time-independent wave equation for a free particle on the circle
is given by
\begin{equation}
\hat{H}\Psi(r,\theta)=-\nabla^{2}\Psi(r,\theta)=\xi\Psi(r,\theta)
\end{equation}
Using the method of separation of variables, we obtain the equivalent
system:
\begin{equation}
\Psi(r,\theta)=R(r)Y(\theta)
\end{equation}
\begin{equation}
\dfrac{d^{2}Y(\theta)}{d\theta^{2}}=-k^{2}Y(\theta)
\end{equation}
\begin{equation}
\dfrac{d^{2}R(r)}{dr^{2}}+\dfrac{1}{r}\dfrac{dR(r)}{dr}+(\dfrac{k^{2}}{r^{2}}-\xi)R(r)=0
\end{equation}
These equations have the well-known solutions of complex exponentials
and Bessel functions. The angular part obeys periodic boundary conditions,
and can be separated into an odd and even part, being the sine and
cosine:
\begin{equation}
Y(\theta)=Ae^{ik\theta}+Be^{-ik\theta}
\end{equation}
\begin{equation}
Y_{\pm}(\theta)=\pm Y_{\pm}(-\theta)
\end{equation}
\begin{equation}
Y(2\pi)=Y(0)
\end{equation}
finding a particular quantisation for the system as given by:
\begin{equation}
k=\pm n
\end{equation}
Rewriting the radial wave equation in natural units $\xi=1$, we find:
\begin{equation}
r^{2}\dfrac{d^{2}R_{n}}{dr^{2}}+r\dfrac{dR_{n}}{dr}+(n^{2}-r^{2})R_{n}=0
\end{equation}
Our solution is then the Bessel function:
\begin{equation}
R_{n}(r)=J_{n}(r)
\end{equation}
This has an integral expansion and inner product:
\begin{equation}
J_{n}(r)=\dfrac{1}{2\pi}\int_{-\pi}^{+\pi}e^{-i(n\phi-r\sin\phi)}d\phi
\end{equation}
\begin{equation}
\int_{-\pi}^{+\pi}J_{m}^{*}(\nu)J_{n}(\nu)d\nu=\dfrac{sin(\pi(m-n))}{2\pi^{2}(m-n)}=\dfrac{\delta(m-n)\delta_{mn}}{2\pi^{2}}
\end{equation}

\section{Spin Waves In Crystals}

Consider a spin wave that moves in a one dimensional crystal, perhaps
along a line. This is the simplest of the continuous degrees of freedom
that can be analysed; we examine it purely to explain the context
of the Bessel function that arose in the previous section, to illustrate
the direct physics of the phenomena in quantum systems that this paper
is concerned with.We take the Hamiltonian:

\begin{equation}
\hat{H}\left|n\right\rangle =-A(\left|n+1\right\rangle +\left|n-1\right\rangle -2\left|n\right\rangle )
\end{equation}
Defining a matrix representation:
\begin{equation}
H_{m,n}=\left\langle n\right|\hat{H}\left|m\right\rangle 
\end{equation}
\begin{equation}
i\dfrac{dC_{n}}{dt}=\sum H_{n,m}C_{m}
\end{equation}
We solve this differential equation with the ansatz:
\begin{equation}
C_{m}(t)=a_{m}e^{-iEt}
\end{equation}
\begin{equation}
Ea_{n}=A(2a_{n}-a_{n-1}-a_{n+1})
\end{equation}
\begin{equation}
\Delta E(p)=E-E_{0}=2\cos(p)
\end{equation}
This eigenvalue equation has a solution that can be written in integral
form as a Green's function:
\begin{equation}
K(\Delta q,t|0,0)=\dfrac{1}{2\pi}\int_{-\pi}^{+\pi}exp(-i(p\Delta q-2t\cos p)).dp
\end{equation}
\begin{equation}
K(\Delta q,t|\mathbf{0})=(-i)^{\Delta q}J_{\Delta q}(2t)
\end{equation}
This is the Bessel function, as required.

\section{Time Dependent Oscillator Revisited}

Take the wave equation in one space dimension, coupled to a time dependent
oscillator:

\begin{equation}
i\hbar\dfrac{\partial\Psi}{\partial t}=-\dfrac{\hbar^{2}}{2m_{e}}\dfrac{\partial^{2}\Psi}{\partial x^{2}}+V_{0}\cos\omega t.\Psi
\end{equation}
Using the cosine transform, and the method of separation of variables:
\begin{equation}
\Psi(x,t)=\phi(x)\beta(t)
\end{equation}
\begin{equation}
z=\cos\omega t\rightleftarrows t=\dfrac{1}{\omega}\cos^{-1}(z)
\end{equation}
we find the differential equation of state:
\begin{equation}
i\dfrac{d\beta}{\beta}=\dfrac{(1-\alpha z)}{\sqrt{1-z^{2}}}dz
\end{equation}
which has an explicit solution:
\begin{equation}
i(\ln\beta(z)-\ln\beta_{0})=-\cos^{-1}(z)+\alpha\sqrt{1-z^{2}}
\end{equation}
This is mid-way between a Bessel function and a Chebyshev polynomial.
Writing $\chi=\omega t$, $z=\cos\chi$we may show a neat set of transformations:
\begin{equation}
\dfrac{\partial\Psi}{\partial\chi}=-\sqrt{1-z^{2}}\dfrac{\partial\Psi}{\partial z}
\end{equation}
\begin{equation}
\dfrac{\partial\Psi}{\partial z}=\dfrac{-1}{\sqrt{1-z^{2}}}\dfrac{\partial\Psi}{\partial\chi}=\dfrac{-1}{\sin\chi}\dfrac{\partial\Psi}{\partial\chi}
\end{equation}
Differentiating these expressions directly, we arrive at two equivalent
differential equations:
\begin{equation}
\dfrac{\partial^{2}\Psi}{\partial\chi^{2}}=\sqrt{1-z^{2}}\dfrac{\partial}{\partial z}(\sqrt{1-z^{2}}\Psi)
\end{equation}
\begin{equation}
\dfrac{\partial^{2}\Psi}{\partial z^{2}}=\dfrac{1}{\sin\chi}\dfrac{\partial}{\partial\chi}(\dfrac{1}{\sin\chi}\dfrac{\partial\Psi}{\partial\chi})
\end{equation}
Expanding these, we find the pair of second order differential equations,
related by the transform pair: 
\begin{equation}
\dfrac{\partial^{2}\Psi}{\partial\chi^{2}}=(1-z^{2})\dfrac{\partial^{2}\Psi}{\partial z^{2}}-z\dfrac{\partial\Psi}{\partial z}
\end{equation}
\begin{equation}
(1-z^{2})\dfrac{\partial^{2}\Psi}{\partial z^{2}}=\dfrac{\partial^{2}\Psi}{\partial\chi^{2}}-\cot\chi\dfrac{\partial\Psi}{\partial\chi}
\end{equation}
\begin{equation}
z\dfrac{\partial\Psi}{\partial z}=\cot\chi\dfrac{\partial\Psi}{\partial\chi}
\end{equation}
Taking the equation of state and directly differentiating we find
a second order differential equation:
\begin{equation}
i(1-z^{2})\dfrac{d^{2}\beta}{dz^{2}}=\beta(-i(1-\alpha z)^{2}-\alpha\sqrt{1-z^{2}}+\dfrac{z(1-\alpha z)}{\sqrt{1-z^{2}}})
\end{equation}
This may be rewritten in the equivalent form:
\begin{equation}
(1-z^{2})\dfrac{d^{2}\beta}{dz^{2}}+(\dfrac{\alpha(1-z^{2})}{1-\alpha z}-z)\dfrac{d\beta}{dz}+(1-\alpha z)^{2}\beta=0
\end{equation}
\begin{equation}
\beta(z)=\beta_{0}exp(i\cos^{-1}(z))exp(-i\alpha\sqrt{1-z^{2}})
\end{equation}

\section{Discrete Fourier Transform For SU(3)}

The discrete Fourier transform (DFT) on a qutrit has subtleties that
are not apparent in the even dimensional case, as there are several
transformations which have similar properties to the expected form.
As the structure of the entire unitary space can be broken down into
three sets of orthogonal rotations and two diagonal operators, defining
exactly what is meant by 'odd' and 'even' is not quite so simple.
This is due to the underlying group structure, as we may define a
third set of properties, namely neutral. We would expect to have three
subgroups which map in a similar way to the DFT on even dimensions,
and one total unitary; this is indeed the case. To show that this
is true, we first generate a cube root of unity:

\begin{equation}
z=\dfrac{1}{2}(-1+i\sqrt{3})
\end{equation}
with complex conjugate:

\begin{equation}
z^{*}=\dfrac{1}{2}(-1-i\sqrt{3})
\end{equation}
This has unit modulus and is a group under $\{*,^{2}\}$:
\begin{equation}
\left|z\right|^{2}=1
\end{equation}
\begin{equation}
z^{2}=z^{*},z=(z^{*})^{2}
\end{equation}
Defining our first DFT matrix: 
\begin{equation}
\hat{R}=\dfrac{1}{\sqrt{3}}\left[\begin{array}{ccc}
1 & 1 & 1\\
1 & z & z^{*}\\
1 & z^{*} & z
\end{array}\right]\approxeq\dfrac{1}{\sqrt{3}}\left[\begin{array}{ccc}
1 & 1 & 1\\
1 & z & z^{2}\\
1 & z^{2} & z^{4}
\end{array}\right]
\end{equation}
This matrix is unitary:
\begin{equation}
\hat{R}\hat{R}^{\dagger}=\hat{R}^{\dagger}\hat{R}=\hat{\mathbf{1}}
\end{equation}
and is a fourth-root of unity$\hat{R}^{4}=\hat{1}$. One useful relationship
is:
\begin{equation}
\hat{R}^{T}\hat{R}=\hat{R}\hat{R}^{T}=\left[\begin{array}{ccc}
1 & 0 & 0\\
0 & 0 & 1\\
0 & 1 & 0
\end{array}\right]
\end{equation}
which is the equivalent C-NOT for qutrits. We may write our complex
unit as the cube root of unity:
\begin{equation}
z^{3}=1
\end{equation}
with equivalent Euler representation:
\begin{equation}
z=exp(\dfrac{2\pi i}{3})
\end{equation}
Let us now consider a particular decomposition of the Hamiltonian:
\begin{equation}
\hat{H}=\left[\begin{array}{ccc}
\omega_{1} & 0 & 0\\
0 & \omega_{2} & 0\\
0 & 0 & \omega_{3}
\end{array}\right]+\hat{Q}\left[\begin{array}{ccc}
1 & 0 & 0\\
0 & -1 & 0\\
0 & 0 & 0
\end{array}\right]\hat{Q^{\dagger}}=\tilde{H}_{0}+\tilde{H}_{Q}
\end{equation}
We define an odd-valued split DFT via:

\begin{equation}
\hat{Q}=\left[\begin{array}{ccc}
1 & 1 & 1\\
1 & k & k^{3}\\
1 & k^{3} & k^{5}
\end{array}\right],\hat{Q^{\dagger}}=\left[\begin{array}{ccc}
1 & 1 & 1\\
1 & k^{-1} & k^{-3}\\
1 & k^{-3} & k^{-5}
\end{array}\right]
\end{equation}
with function values:

\begin{equation}
k=e^{i\theta},k^{-1}=k^{*}
\end{equation}
This takes our Hamiltonian to the operator:
\begin{equation}
\hat{H}=\left[\begin{array}{ccc}
\omega_{1} & 1-k^{-1} & 1-k^{-3}\\
1-k & \omega_{2} & 1-k^{-2}\\
1-k^{3} & 1-k^{2} & \omega_{3}
\end{array}\right]
\end{equation}
Writing out the formulae for the trigonometric polynomials: 
\begin{equation}
k^{3}=\cos\theta(4\cos^{2}\theta-3)+i\sin\theta(4\sin^{2}\theta-3)
\end{equation}
\begin{equation}
k^{2}=2cos^{2}\theta-1+i2sin\theta cos\theta
\end{equation}
\begin{equation}
k^{-n}=(k^{n})^{*}
\end{equation}
Note that this takes the Hamiltonian to the off-diagonal non-comutative
submanifold; consequently we expect to find two equivalent different
split DFTs which take the Hamiltonian to the same type of submanifolds.
This type of transformation will not be unitary or invertible, it
is more along the same lines as making a projection onto a co-ordinate
axis, keeping a subset of the co-ordinates and discarding the rest
of the set. Writing out the transformation law on the Hamiltonian
matrix:
\begin{equation}
\hat{H}_{W}=\hat{W}\left[\begin{array}{ccc}
1 & 0 & 0\\
0 & -1 & 0\\
0 & 0 & 0
\end{array}\right]\hat{W^{\dagger}}
\end{equation}
We now substitute the even-valued split DFT into this relation:
\begin{equation}
\hat{W}=\left[\begin{array}{ccc}
1 & 1 & 1\\
1 & w & w^{2}\\
1 & w^{2} & w^{4}
\end{array}\right]
\end{equation}
and find the explicitly transformed matrix:
\begin{equation}
\hat{H}_{W}=\left[\begin{array}{ccc}
0 & 1-w^{-1} & 1-w^{-2}\\
1-w & 0 & 1-w^{-1}\\
1-w^{2} & 1-w & 0
\end{array}\right]
\end{equation}
\begin{equation}
w=e^{i\chi}
\end{equation}
This matrix has two independent component functions, thereby satisfying
the physical conditions of the SU(3)/SU(2) submanifold. We may write
this in the form:
\begin{equation}
\tilde{H}_{W}=\hat{A}+\hat{A}^{T}+\hat{X}\cos\chi+\hat{Y}\sin\chi+\hat{K}+\hat{K}^{\dagger}
\end{equation}
Consider the operators $\hat{X}\hat{,Y}$: 
\begin{equation}
\hat{X}=\left[\begin{array}{ccc}
0 & 1 & 0\\
1 & 0 & 1\\
0 & 1 & 0
\end{array}\right],\hat{Y}=\left[\begin{array}{ccc}
0 & -i & 0\\
i & 0 & -i\\
0 & i & 0
\end{array}\right]
\end{equation}
They form a non-commutative subgroup that is similar to SU(2):
\begin{equation}
[\hat{X},\hat{Y}]=2i\left[\begin{array}{ccc}
1 & 0 & 0\\
0 & 0 & 0\\
0 & 0 & -1
\end{array}\right]=2i\hat{Z}
\end{equation}
\begin{equation}
[\hat{X},\hat{Z}]=-i\hat{Y}
\end{equation}
\begin{equation}
[\hat{Y},\hat{Z}]=i\hat{X}
\end{equation}
Up to a constant matrix, this Hamiltonian is equivalent to the SO(3)
subalgebra of SU(3). The constant matrix is the set of upper and lower
projectors, with representation:
\begin{equation}
\hat{A}=\hat{A}_{1}+\hat{A}_{2}+\hat{A}_{3}
\end{equation}

\begin{equation}
\hat{A}_{1}=\left|1\left\rangle \right\langle 2\right|,\hat{A}_{2}=\left|1\left\rangle \right\langle 3\right|,\hat{A}_{3}=\left|2\left\rangle \right\langle 3\right|
\end{equation}

\begin{equation}
\hat{A}_{j}^{\dagger}=\hat{A}_{j}^{T}
\end{equation}
This matrix is not unique, one may define another matrix, given by:

\begin{equation}
\hat{J}=\left[\begin{array}{ccc}
1 & 1 & 1\\
1 & j & j^{2}\\
1 & j^{2} & j^{3}
\end{array}\right]
\end{equation}
Transforming the Hamiltonian we find:
\begin{equation}
\hat{H}_{J}=\hat{J}\left[\begin{array}{ccc}
1 & 0 & 0\\
0 & -1 & 0\\
0 & 0 & 0
\end{array}\right]\hat{J^{\dagger}}
\end{equation}
\begin{equation}
\hat{H}_{J}=\left[\begin{array}{ccc}
0 & 1-j^{-1} & 1-j^{-2}\\
1-j & 0 & 1-j^{-1}\\
1-j^{2} & 1-j & 0
\end{array}\right]
\end{equation}
which means our mapped Hamiltonian is an equivalent two-parameter
subgroup. Writing this as a congruence relation:
\begin{equation}
\hat{H}_{Q}\ncong\hat{H}_{J}\cong\hat{H}_{W}
\end{equation}
The functions involved have decompositions that may be readily calculated
with De-Moivre's theorem:
\begin{equation}
1-j^{2}=1-e^{2i\theta}=1-\cos2\theta-i\sin2\theta
\end{equation}
\begin{equation}
1-\cos2\theta-i\sin2\theta=2(\cos^{2}\theta-1)-i2\sin\theta\cos\theta
\end{equation}
\begin{equation}
1-j^{2}=2(\sin^{2}\theta)-i2\sin\theta\cos\theta
\end{equation}
These functions have the necessary symmetry under conjugation to give
us a Hermitian matrix:
\begin{equation}
j^{-n}=(j^{n})^{*}
\end{equation}
\begin{equation}
1-j^{-l}=(1-j^{l})^{*}
\end{equation}
None of the split-transforms $\hat{J},\hat{Q}$or $\hat{W}$ are unitary
or orthogonal; the only matrix which has the required property to
be a DFT in its own right is $\hat{R}$.

\section{Dihedral Group On The Unit Triangle}

The permutations one may apply to a triangle with uniquely labelled
vertices generate a dihedral group, defined by the matrices:

\begin{equation}
\hat{S}_{1}=\left[\begin{array}{ccc}
1 & 0 & 0\\
0 & 1 & 0\\
0 & 0 & 1
\end{array}\right],\hat{S}_{2}=\left[\begin{array}{ccc}
0 & 0 & 1\\
0 & 1 & 0\\
1 & 0 & 0
\end{array}\right]
\end{equation}
\begin{equation}
\hat{S}_{3}=\left[\begin{array}{ccc}
1 & 0 & 0\\
0 & 0 & 1\\
0 & 1 & 0
\end{array}\right],\hat{S}_{4}=\left[\begin{array}{ccc}
0 & 1 & 0\\
1 & 0 & 0\\
0 & 0 & 1
\end{array}\right]
\end{equation}
\begin{equation}
\hat{S}_{5}=\left[\begin{array}{ccc}
0 & 0 & 1\\
1 & 0 & 0\\
0 & 1 & 0
\end{array}\right],\hat{S}_{6}=\left[\begin{array}{ccc}
0 & 1 & 0\\
0 & 0 & 1\\
1 & 0 & 0
\end{array}\right]
\end{equation}
This is the smallest non-commutative group. Note that these are a
group of rotoflections, as opposed to proper reflections or rotation
matrices. Using our Hamiltonian operators, we may define a closely
related series of operators using our initial value equations:

\begin{equation}
\hat{Q}(0)=\left[\begin{array}{ccc}
\dfrac{1}{\sqrt{2}} & -\dfrac{1}{\sqrt{2}} & 0\\
\dfrac{1}{\sqrt{2}} & \dfrac{1}{\sqrt{2}} & 0\\
0 & 0 & 1
\end{array}\right]
\end{equation}
This is one of our equivalent Hadamard gates, which constitutes a
pure rotation in 3-D space.
\begin{equation}
\hat{Q}^{2}(0)=\left[\begin{array}{ccc}
0 & -1 & 0\\
1 & 0 & 0\\
0 & 0 & 1
\end{array}\right]
\end{equation}
We then have a reflection in a plane of symmetry, which is a signed
permutation matrix.
\begin{equation}
\hat{D}(0)=\left[\begin{array}{ccc}
-\dfrac{i}{\sqrt{2}} & \dfrac{i}{\sqrt{2}} & 0\\
\dfrac{1}{\sqrt{2}} & \dfrac{1}{\sqrt{2}} & 0\\
0 & 0 & i
\end{array}\right]
\end{equation}
Because we are working on higher dimensions, there is a greater number
of available operators, which we need in order to close the algebra
and conduct meaningful calculations. With qubits all that is required
is a phase gate, a NOT gate, a Hadamard matrix and a XOR gate.
\begin{equation}
\hat{D}^{2}(0)=\frac{1}{2}\left[\begin{array}{ccc}
i-1 & i+1 & 0\\
-i+1 & i+1 & 0\\
0 & 0 & -2
\end{array}\right]
\end{equation}
\begin{equation}
\hat{J}(0)=\left[\begin{array}{ccc}
\dfrac{1}{\sqrt{2}} & -\dfrac{1}{\sqrt{2}} & 0\\
\dfrac{1}{\sqrt{2}} & \dfrac{1}{\sqrt{2}} & 0\\
0 & 0 & i
\end{array}\right]
\end{equation}
\begin{equation}
\hat{J}^{4}(0)=\left[\begin{array}{ccc}
-1 & 0 & 0\\
0 & -1 & 0\\
0 & 0 & 1
\end{array}\right]
\end{equation}
The particular matrices we have been considering above play the role
of the phase gate, one of the possible equivalent C-NOT trit gates
and the Hadamard gate in time optimal qutrit quantum computation.
They are signed complex permutation matrices. The extremals of the
unitary generators are at the points $\{0,\dfrac{\pi}{2},\pi,\dfrac{3\pi}{2}\}$
which define a base set of qutrit computational gates:

\begin{equation}
\hat{Q}(0)=\left[\begin{array}{ccc}
\dfrac{1}{\sqrt{2}} & -\dfrac{1}{\sqrt{2}} & 0\\
\dfrac{1}{\sqrt{2}} & \dfrac{1}{\sqrt{2}} & 0\\
0 & 0 & 1
\end{array}\right],\hat{Q}(\pi)=\left[\begin{array}{ccc}
-\dfrac{1}{\sqrt{2}} & \dfrac{1}{\sqrt{2}} & 0\\
\dfrac{1}{\sqrt{2}} & \dfrac{1}{\sqrt{2}} & 0\\
0 & 0 & -1
\end{array}\right]
\end{equation}
\begin{equation}
\hat{Q}(\dfrac{\pi}{2})=\left[\begin{array}{ccc}
0 & 0 & ie^{-i\varrho}\\
\dfrac{1}{\sqrt{2}} & \dfrac{1}{\sqrt{2}} & 0\\
\dfrac{ie^{i\varrho}}{\sqrt{2}} & \dfrac{-ie^{i\varrho}}{\sqrt{2}} & 0
\end{array}\right]
\end{equation}

\begin{equation}
\hat{Q}(\dfrac{3\pi}{2})=\left[\begin{array}{ccc}
0 & 0 & -ie^{-i\varrho}\\
\dfrac{1}{\sqrt{2}} & \dfrac{1}{\sqrt{2}} & 0\\
-\dfrac{ie^{i\varrho}}{\sqrt{2}} & \dfrac{ie^{i\varrho}}{\sqrt{2}} & 0
\end{array}\right]
\end{equation}
It is possible to classify these matrices by their eigenvalues and
determinants into various point symmetry groups. 

\noindent In some sense the unitary matrices we are using play the
role of {}``square roots'' of unitary matrices themselves, which
are in fact the dihedral group above. Being the complete set of positive
and negative roots of the group expression, we receive twelve unique
operators which may be readily used to generate a wide variety of
useful operations. 
\begin{equation}
\hat{J}(0)=\left[\begin{array}{ccc}
\dfrac{1}{\sqrt{2}} & -\dfrac{1}{\sqrt{2}} & 0\\
\dfrac{1}{\sqrt{2}} & \dfrac{1}{\sqrt{2}} & 0\\
0 & 0 & i
\end{array}\right],\hat{J}(\pi)=\left[\begin{array}{ccc}
-\dfrac{1}{\sqrt{2}} & \dfrac{1}{\sqrt{2}} & 0\\
\dfrac{1}{\sqrt{2}} & \dfrac{1}{\sqrt{2}} & 0\\
0 & 0 & -i
\end{array}\right]
\end{equation}
\begin{equation}
\hat{J}(\dfrac{\pi}{2})=\left[\begin{array}{ccc}
0 & 0 & -1\\
\dfrac{1}{\sqrt{2}} & \dfrac{1}{\sqrt{2}} & 0\\
\dfrac{i}{\sqrt{2}} & \dfrac{-i}{\sqrt{2}} & 0
\end{array}\right],\hat{J}(\dfrac{3\pi}{2})=\left[\begin{array}{ccc}
0 & 0 & 1\\
\dfrac{1}{\sqrt{2}} & \dfrac{1}{\sqrt{2}} & 0\\
-\dfrac{i}{\sqrt{2}} & \dfrac{i}{\sqrt{2}} & 0
\end{array}\right]
\end{equation}
\begin{equation}
\hat{D}(0)=\left[\begin{array}{ccc}
-\dfrac{i}{\sqrt{2}} & \dfrac{i}{\sqrt{2}} & 0\\
\dfrac{1}{\sqrt{2}} & \dfrac{1}{\sqrt{2}} & 0\\
0 & 0 & i
\end{array}\right],\hat{D}(\pi)=\left[\begin{array}{ccc}
\dfrac{i}{\sqrt{2}} & -\dfrac{i}{\sqrt{2}} & 0\\
\dfrac{1}{\sqrt{2}} & \dfrac{1}{\sqrt{2}} & 0\\
0 & 0 & -i
\end{array}\right]
\end{equation}
\begin{equation}
\hat{D}(\dfrac{\pi}{2})=\left[\begin{array}{ccc}
0 & 0 & i\\
\dfrac{1}{\sqrt{2}} & \dfrac{1}{\sqrt{2}} & 0\\
\dfrac{i}{\sqrt{2}} & \dfrac{-i}{\sqrt{2}} & 0
\end{array}\right],\hat{D}(\dfrac{3\pi}{2})=\left[\begin{array}{ccc}
0 & 0 & -i\\
\dfrac{1}{\sqrt{2}} & \dfrac{1}{\sqrt{2}} & 0\\
-\dfrac{i}{\sqrt{2}} & \dfrac{i}{\sqrt{2}} & 0
\end{array}\right]
\end{equation}
This set of unitary transformations is necessary and sufficient to
implement any quantum trinary computation. Some elementary operations
include:
\begin{equation}
\hat{D}(0)\hat{D}(\dfrac{\pi}{2})-\hat{D}(\dfrac{\pi}{2})\hat{D}(0)=\left[\begin{array}{ccc}
i & i & 0\\
i & -i & 0\\
0 & 0 & 0
\end{array}\right]
\end{equation}

\begin{equation}
\hat{D}(0)\hat{D}(\dfrac{\pi}{2})+\hat{D}(\dfrac{\pi}{2})\hat{D}(0)=\left[\begin{array}{ccc}
1 & -1 & 0\\
1 & 1 & 0\\
0 & 0 & 2
\end{array}\right]
\end{equation}
Other operators will be explored in future works. By classification
of the eigenvalues, these matrices may be defined as a certain types
of point and continuous symmetry groups in the complex plane, which
are related to the roots of polynomial equations which we explored
earlier. This method of qutrit computation and modelling of quantum
states and control is efficient, in that it is time optimal; resource
bounded, in that there is an existing physical overhead on the total
energy imparted to the system, reversible, in that it is unitary,
and robust to error, as it is periodic. We could consider our SU(3)
system to be in all ways equivalent to the physical sum of a qubit
and an ancilla. In this sense, we are moving beyond qubit quantum
computation into new realms of possibility. With a qubit and an ancilla
it is possible to achieve much more than with a single qubit, or any
number of coupled qubits. Data correction requires ancillas, and the
nature of quantum states is that all indistinguishable alternatives
are to be summed over; this necessitates our examination of qutrits
both as quantum computational objects in themselves, and as possible
sources of error within a qubit calculation. This is the simplest
example of a odd dimensional quantum state; in this particularly simple
case we are fortunately able to derive rich symmetries which are descriptive
of many of the full properties of groups which do not emerge in lower
dimensional cases.

\section{Subgroups of SU(3)}

Explicitly writing out the diagonal generators of SU(3), we find:

\begin{equation}
\frac{1}{\sqrt{3}}\left[\begin{array}{ccc}
\gamma & 0 & 0\\
0 & \gamma & 0\\
0 & 0 & -2\gamma
\end{array}\right],\left[\begin{array}{ccc}
\vartheta & 0 & 0\\
0 & -\vartheta & 0\\
0 & 0 & 0
\end{array}\right]\backsim\left[\begin{array}{ccc}
\omega_{1} & 0 & 0\\
0 & \omega_{2} & 0\\
0 & 0 & \omega_{3}
\end{array}\right]
\end{equation}
The diagonal subgroup has one component which commutes with every
other member of the group, the other only commutes with the rotation
subgroup. We may therefore use the diagonal component to define the
group structure, as it does not matter whether the commuting operator
is taken as part of the constraint or Hamiltonian. The matrix structure
of the possible subdivisions is then:

\begin{equation}
\left[\begin{array}{ccc}
\omega_{1} & 0 & 0\\
0 & \omega_{2} & 0\\
0 & 0 & \omega_{3}
\end{array}\right],\left[\begin{array}{ccc}
0 & \epsilon_{1} & \epsilon_{2}\\
\epsilon_{1}^{*} & 0 & \epsilon_{3}\\
\epsilon_{2}^{*} & \epsilon_{3}^{*} & 0
\end{array}\right]
\end{equation}

\begin{equation}
\left[\begin{array}{ccc}
\omega_{1} & 0 & \kappa\\
0 & \omega_{2} & 0\\
\kappa^{*} & 0 & \omega_{3}
\end{array}\right],\left[\begin{array}{ccc}
0 & \epsilon_{1} & 0\\
\epsilon_{1}^{*} & 0 & \epsilon_{2}\\
0 & \epsilon_{2}^{*} & 0
\end{array}\right]
\end{equation}
\begin{equation}
\left[\begin{array}{ccc}
0 & 0 & \kappa\\
0 & 0 & 0\\
\kappa^{*} & 0 & 0
\end{array}\right],\left[\begin{array}{ccc}
\omega_{1} & \epsilon_{1} & 0\\
\epsilon_{1}^{*} & \omega_{2} & \epsilon_{2}\\
0 & \epsilon_{2}^{*} & \omega_{3}
\end{array}\right]
\end{equation}

\begin{equation}
\left[\begin{array}{ccc}
\vartheta & \beta & 0\\
\beta^{*} & -\vartheta & 0\\
0 & 0 & 0
\end{array}\right],\left[\begin{array}{ccc}
\gamma & 0 & \epsilon_{1}\\
0 & \gamma & \epsilon_{2}\\
\epsilon_{1}^{*} & \epsilon_{2}^{*} & -2\gamma
\end{array}\right],
\end{equation}
For any choice of Hamiltonian operator and associated constraint,
we either find one or the other to be periodic (or constant). The
equations of dynamic state are equivalent up to isometry with a permutation
matrix.

\section{Geodesic On SU(4)}

Let us examine a system where we can only apply global control pulses,
without individual addressing, on two qubits. The Hamiltonian matrix
is the Heisenberg model:

\begin{equation}
\tilde{H}[t]=\lambda_{x}\hat{\sigma}_{x}\otimes\hat{\sigma}_{x}+\lambda_{y}\hat{\sigma}_{y}\otimes\hat{\sigma}_{y}+\lambda_{z}\hat{\sigma}_{z}\otimes\hat{\sigma}_{z}
\end{equation}

\noindent In matrix form: 
\begin{equation}
\tilde{H}[t]=\left[\begin{array}{cccc}
\lambda_{z} & 0 & 0 & \lambda_{-}\\
0 & -\lambda_{z} & \lambda_{+} & 0\\
0 & \lambda_{+} & -\lambda_{z} & 0\\
\lambda_{-} & 0 & 0 & \lambda_{z}
\end{array}\right];\lambda_{\pm}=\lambda_{x}\pm\lambda_{y}
\end{equation}
The constraint may be expanded in the Hermitian basis:

\noindent 
\begin{equation}
\tilde{F}[t]=(\vec{m}(t)\cdot\vec{\sigma})\otimes\hat{\mathbf{1}}+\hat{\mathbf{1}}\otimes(\vec{n}(t)\cdot\vec{\sigma})+\sum_{i\neq j}\Xi_{i,j}(t)\hat{\sigma}_{i}\otimes\hat{\sigma}_{j}
\end{equation}

\noindent Evaluating the quantum brachistochrone, we find the vector
relationship:
\begin{equation}
\dfrac{d}{dt}\left[\begin{array}{c}
\lambda_{x}\\
\lambda_{y}\\
\lambda_{z}
\end{array}\right]=0
\end{equation}

\noindent Our time optimal Hamiltonian is therefore a constant matrix,
and may be exponentiated directly. Our wave-vector after some time
is given by:
\begin{equation}
\left|\psi(t)\right\rangle =\exp(-i\tilde{H}t)\left|\psi(0)\right\rangle 
\end{equation}

\noindent Evaluating the quantum brachistochrone equation, and choosing
the initial state $\left|\psi(0)\right\rangle =[1,0,0,0]^{T}$ we
immediately find $\lambda_{x}=-\lambda_{y}$. Our final state is:
\begin{equation}
\left|\psi(T)\right\rangle =\frac{1}{\sqrt{2}}(\left|00\right\rangle -i\left|11\right\rangle )
\end{equation}

\noindent from which we obtain $\lambda_{z}=0$. Our state evolves
according to:
\begin{equation}
\left|\psi(t)\right\rangle =\frac{1}{\sqrt{2}}\left[\begin{array}{c}
\mathrm{cos}(2\lambda_{x}t)\\
0\\
0\\
-i\mathrm{sin}(2\lambda_{x}t)
\end{array}\right]
\end{equation}
and therefore our energy-time relationship is $T=\dfrac{\pi}{\lambda_{x}}$.

\section{Dirac Equation in Co-rotating Frame}

We take the more general ansatz for a Hamiltonian on SU(4):

\begin{equation}
\tilde{H}=\left[\begin{array}{cccc}
\alpha & 0 & p_{z} & \varepsilon\\
0 & \alpha & \varepsilon^{*} & -p_{z}\\
p_{z} & \varepsilon & -\alpha & 0\\
\varepsilon^{*} & -p_{z} & 0 & -\alpha
\end{array}\right]=\left[\begin{array}{cc}
\alpha\hat{1} & -i\vec{\beta}\cdot\vec{\sigma}\\
i\vec{\beta}\cdot\vec{\sigma} & -\alpha\hat{1}
\end{array}\right]
\end{equation}
Evaluating the determinant:
\begin{equation}
det(\tilde{H}-\lambda\hat{1})=0
\end{equation}
Obtaining eigenvalue equation:
\begin{equation}
(-\lambda^{2}+\alpha^{2}+p_{z}^{2}+\left|\varepsilon\right|^{2})^{2}=0
\end{equation}
\begin{equation}
\lambda=\pm\sqrt{\alpha^{2}+p_{z}^{2}+\left|\varepsilon\right|^{2}}
\end{equation}
The isotropic constraint gives:
\begin{equation}
Tr(\dfrac{\tilde{H}^{2}}{2})=\mathrm{const.}=\alpha^{2}+p_{z}^{2}+\left|\varepsilon\right|^{2}
\end{equation}
We therefore rescale the time parameter such that $\alpha^{2}+p_{z}^{2}+\left|\varepsilon\right|^{2}=1$,
so our original Hamiltonian matrix has the form of a 4-dimensional
unit vector. The Hamiltonian matrix is difficult to work with, in
strict terms of evaluating the quantum brachistochrone equation; the
solutions are not obvious, and the bilinearity of the differential
equations causes difficulty in finding an exact solution. We therefore
must use a unitary transformation, equivalent to the Hadamard gate
on the Pauli matrices, that acts on the 4-spinor. The necessary transformation
is given by: 
\begin{equation}
\hat{W}=\dfrac{1}{\sqrt{2}}\left[\begin{array}{cc}
\hat{1} & \hat{1}\\
\hat{1} & -\hat{1}
\end{array}\right]=\dfrac{1}{\sqrt{2}}(\hat{\sigma}_{x}+\hat{\sigma}_{z})\otimes\hat{1}
\end{equation}
\begin{equation}
\hat{W}\hat{W}^{\dagger}=\hat{W}^{\dagger}\hat{W}=\mathbf{1}
\end{equation}
\begin{equation}
\hat{W}=\hat{W}^{\dagger}
\end{equation}
Using the standard symmetry transformation, we obtain a new Hamiltonian
matrix: 
\begin{equation}
\tilde{H}_{W}=\hat{W}\tilde{H}\hat{W}^{\dagger}=\left[\begin{array}{cc}
\hat{0} & \hat{b}\\
\hat{b}^{\dagger} & \hat{0}
\end{array}\right]
\end{equation}
\begin{equation}
\hat{b}=\alpha\hat{1}+i\vec{\beta}\cdot\vec{\sigma}
\end{equation}
\begin{equation}
\hat{b}^{\dagger}=\alpha\hat{1}-i\vec{\beta}\cdot\vec{\sigma}
\end{equation}
We then form matrices of eigenvectors:
\begin{equation}
\hat{P}=\left[\begin{array}{cc}
\hat{1} & \hat{1}\\
\hat{b} & -\hat{b}
\end{array}\right]
\end{equation}
\begin{equation}
\hat{P}^{\dagger}=\left[\begin{array}{cc}
\hat{1} & \hat{b}^{\dagger}\\
\hat{1} & -\hat{b}^{\dagger}
\end{array}\right]
\end{equation}
\begin{equation}
\hat{P}\hat{P}^{\dagger}=\hat{P}^{\dagger}\hat{P}=\mathbf{1}
\end{equation}
\begin{equation}
\tilde{H}_{WP}=\hat{P}\tilde{H}_{W}\hat{P}^{\dagger}=\left[\begin{array}{cc}
\hat{1} & 0\\
0 & -\hat{1}
\end{array}\right]=\hat{\sigma}_{z}\otimes\hat{1}
\end{equation}
The time evolution operator for the original system may be written
as:
\begin{equation}
\hat{U}(t,0)=exp(-i\int_{0}^{t}\tilde{H}(s)ds)
\end{equation}
In the transformed reference frame this reads as:
\begin{equation}
\hat{U}(t,0)=exp(-i\int_{0}^{t}\hat{Q}^{\dagger}(s)\tilde{H}_{WP}\hat{Q}(s)ds)
\end{equation}
The unitary operator which transforms the Hamiltonian is of the form
$\hat{Q}(s)=\hat{P}(s)\hat{W}$. This has the useful property:
\begin{equation}
[\tilde{H}_{WP},\hat{Q}(s)]=0
\end{equation}
which enables the direct evaluation of the time evolution operator:
\begin{equation}
\hat{U}(t,0)=\left[\begin{array}{cc}
\hat{1}e^{it} & 0\\
0 & \hat{1}e^{-it}
\end{array}\right]
\end{equation}
Using this unitary to transform the initial condition of the Hamiltonian
matrix, we find:
\begin{equation}
\tilde{H}(t)=\hat{U}(t,0)\tilde{H}(0)\hat{U}^{\dagger}(t,0)=\tilde{H}_{0}+\tilde{V}[t,\vec{\beta}(0)]
\end{equation}
\begin{equation}
\tilde{H}(t)=\left[\begin{array}{cc}
\alpha(0)\hat{1} & e^{-2it}\hat{A}^{\dagger}(0)\\
e^{2it}\hat{A}(0) & -\alpha(0)\hat{1}
\end{array}\right]
\end{equation}
\begin{equation}
\tilde{H}(t+T)=\tilde{H}(t)
\end{equation}
\begin{equation}
det(\tilde{H}(t)-\lambda\hat{1})=det(\tilde{H}(0)-\lambda\hat{1})=0
\end{equation}
The matrix constraint for the time optimal control problem can then
be written as:
\begin{equation}
\tilde{F}(t)=\left[\begin{array}{cc}
\begin{array}{cc}
0 & \xi_{1}\\
\xi_{1}^{*} & 0
\end{array} & e^{-2it}\hat{A}\\
e^{2it}\hat{A}^{\dagger} & \begin{array}{cc}
0 & \xi_{2}\\
\xi_{2}^{*} & 0
\end{array}
\end{array}\right]
\end{equation}
where $\hat{A}=\left[\begin{array}{cc}
a & b\\
c & d
\end{array}\right]$\,and all the variables involved (other than the obvious time dependence)
are constants. This Hamiltonian is the block-diagonal solution to
the Klein-Gordon equation, in that we may write: 
\begin{equation}
\tilde{H}^{2}=(\alpha^{2}+p_{z}^{2}+\left|\varepsilon\right|^{2})\left[\begin{array}{cc}
\hat{1} & 0\\
0 & \hat{1}
\end{array}\right]
\end{equation}
Obviously this is invariant under the time dependent transformation
on the Hamiltonian. It is possible to map this solution unitarily
to a number of other physical systems which describe the motion of
electrons and positrons in different basis sets. For completeness,
these matrices are listed below.
\begin{equation}
\hat{U}_{3}=\left[\begin{array}{cccc}
0 & 0 & 1 & 0\\
0 & 0 & 0 & 1\\
1 & 0 & 0 & 0\\
0 & 1 & 0 & 0
\end{array}\right]
\end{equation}
\begin{equation}
\hat{U}_{4}=\dfrac{1}{\sqrt{2}}\left[\begin{array}{cccc}
1 & 0 & 1 & 0\\
i & 0 & -i & 0\\
0 & 1 & 0 & 1\\
0 & -i & 0 & i
\end{array}\right]
\end{equation}
\begin{equation}
\hat{U}_{5}=\dfrac{1}{\sqrt{2}}\left[\begin{array}{cccc}
1 & -i & 0 & 0\\
1 & i & 0 & 0\\
0 & 0 & 1 & -i\\
0 & 0 & 1 & i
\end{array}\right]
\end{equation}
\begin{equation}
\hat{U}_{6}=\dfrac{1}{\sqrt{2}}\left[\begin{array}{cccc}
1 & 0 & 1 & 0\\
0 & 1 & 0 & 1\\
1 & 0 & -1 & 0\\
0 & 1 & 0 & -1
\end{array}\right]
\end{equation}
\begin{equation}
\hat{U}_{7}=\dfrac{1}{\sqrt{2}}\left[\begin{array}{cccc}
1 & 0 & 0 & 1\\
0 & 1 & 1 & 0\\
0 & 1 & -1 & 0\\
1 & 0 & 0 & -1
\end{array}\right]
\end{equation}
\begin{equation}
\hat{U}_{8}=\dfrac{1}{\sqrt{2}}\left[\begin{array}{cccc}
1 & 1 & -1 & 1\\
1 & 1 & 1 & -1\\
-1 & 1 & 1 & 1\\
1 & -1 & 1 & 1
\end{array}\right]
\end{equation}
\begin{equation}
\hat{U}_{8}=\dfrac{1}{2}\left[\begin{array}{cccc}
1 & 1 & 1 & 1\\
1 & -i & -1 & i\\
1 & -1 & 1 & -1\\
1 & i & -1 & -i
\end{array}\right]
\end{equation}
\begin{equation}
\hat{U}_{9}=\left[\begin{array}{cccc}
1 & 0 & 0 & 0\\
0 & 0 & 0 & 1\\
0 & 1 & 0 & 0\\
0 & 0 & 1 & 0
\end{array}\right]
\end{equation}
\begin{equation}
\hat{U}_{10}=\left[\begin{array}{cccc}
1 & 0 & 0 & 0\\
0 & 1 & 0 & 0\\
0 & 0 & 0 & 1\\
0 & 0 & 1 & 0
\end{array}\right]
\end{equation}
These transformation matrices mean that, given we have solved the
specific problem of the Dirac matrix, that a large set of problems
fit into the equivalence class, which saves a great deal of effort
developing particular solutions. Note also that the Hamiltonian has
a certain periodicity, which can be written as $T_{min}\times\left\Vert E\right\Vert =\pi$,
this relationship can be extended unitarily to all the other equivalent
representations by isometry.

\section{Dimensional Arrays Of Arbitrary Size}

Take an arbitrary tracefree matrix in $SU(n)$, which we can express
as the array:

\begin{equation}
\tilde{X}=\left[\begin{array}{ccccc}
\omega_{1} & \alpha_{1} & \alpha_{2} & \cdots & \vdots\\
\alpha_{1}^{*} & \omega_{2} & \alpha_{3} & \cdots & \vdots\\
\alpha_{2}^{*} & \alpha_{3}^{*} & \ddots & \ldots & \alpha_{n-1}\\
\vdots & \cdots & \ddots & \ddots & \alpha_{n}\\
\vdots & \cdots & \alpha_{n-1}^{*} & \alpha_{n}^{*} & \omega_{n}
\end{array}\right],\sum_{j}\omega_{j}=0
\end{equation}

\noindent \begin{flushleft}
This can be mapped invertibly and one-to-one to the vector 
\par\end{flushleft}

\noindent \begin{flushleft}
\begin{equation}
\overrightarrow{X}=\left[\begin{array}{ccccccccc}
(\omega_{1}, & \cdots, & \omega_{n}), & (\alpha_{1}, & \cdots, & \alpha_{k}), & (\alpha_{1}^{*}, & \cdots, & \alpha_{k}^{*})\end{array}\right]^{T}
\end{equation}
 which exists in the space $\left[\begin{array}{ccc}
\mathbb{R^{\mathrm{n-1}}}, & \mathbb{C}^{k}, & \mathbb{C}^{k*}\end{array}\right]^{T}$, and the dimension of the complex space is $k=\sum_{1}^{n-1}j$.
One of the components of the real vector within the multivector is
dependent; we retain the dummy variable to aid in the analysis. In
any real calculation we immediately remove the redundant variable
once the differential equations are calculated. If we choose the particular
Hamiltonian matrix: 
\begin{equation}
\widetilde{H}=\left[\begin{array}{cccccc}
\omega_{1} & 0 & \cdots & \cdots & 0 & \varepsilon_{k}\\
0 & \omega_{2} & \cdots & \cdots & \varepsilon_{k-1} & 0\\
\vdots & \vdots & . & . & \vdots & \vdots\\
\vdots & \vdots & . & . & \vdots & \vdots\\
0 & \varepsilon_{k-1}^{*} & \cdots & \cdots & \omega_{n-1} & 0\\
\varepsilon_{k}^{*} & 0 & \cdots & \cdots & 0 & \omega_{n}
\end{array}\right]
\end{equation}
 
\par\end{flushleft}

\begin{equation}
\tilde{H}=\sum_{n}\omega_{n}(t)\left|n\right\rangle \left\langle n\right|+\sum_{k+j=n}(\varepsilon_{j}(t)\left|k\right\rangle \left\langle j\right|+\mathrm{h.c.})
\end{equation}
 Then the constraint is given by: 
\begin{equation}
\tilde{F}=\sum_{i\neq j,i+j\neq n}(\alpha_{i,j}(t)\left|i\right\rangle \left\langle j\right|+\mathrm{h.c.})
\end{equation}
 
\begin{equation}
\widetilde{F}=\left[\begin{array}{cccccc}
0 & \beta_{1,2} & \cdots & \cdots & \beta_{1,n-1} & 0\\
\beta_{1,2}^{*} & 0 & \cdots & \cdots & 0 & \beta_{2,n}\\
\vdots & \vdots & . & . & \vdots & \vdots\\
\vdots & \vdots & . & . & \vdots & \vdots\\
\beta_{1,n-1}^{*} & 0 & \cdots & \cdots & 0 & \beta_{n-1,n}\\
0 & \beta_{2,n}^{*} & \cdots & \cdots & \beta_{n-1,n}^{*} & 0
\end{array}\right]
\end{equation}
 We may write the quantum brachistochrone as the form: 
\begin{equation}
i\frac{d}{dt}(\widetilde{H}[t]+\widetilde{F}[t])\in G_{F}
\end{equation}
 Hence the Hamiltonian is constant:

\begin{equation}
\frac{d}{dt}\widetilde{H}[t]=0
\end{equation}
Note that, by induction, we have shown the existence of a geodesic
on SU(n), having shown the existence of geodesics on SU(2), SU(2+1),
SU(n) and SU(n+1). Another family of infinite dimensional matrices
is given by: 
\begin{equation}
\widetilde{H}=\left[\begin{array}{cccccc}
0 & \varepsilon_{1} & 0 & \cdots & \cdots & 0\\
\varepsilon_{1}^{*} & 0 & \varepsilon_{2} & 0 & \cdots & \vdots\\
0 & \varepsilon_{2}^{*} & \ddots & \ddots & 0 & \vdots\\
\vdots & 0 & \ddots & \ddots & \varepsilon_{n-1} & 0\\
\vdots & \cdots & 0 & \varepsilon_{n-1}^{*} & 0 & \varepsilon_{n}\\
0 & \cdots & \cdots & 0 & \varepsilon_{n}^{*} & 0
\end{array}\right]
\end{equation}
 with associated constraint matrix $\tilde{F}=$ 
\begin{equation}
\left[\begin{array}{ccccccc}
\omega_{1} & 0 & \beta_{1,3} & \beta_{1,4} & \cdots & \beta_{1,n-1} & \beta_{1,n}\\
0 & \omega_{2} & 0 & \beta_{2,4} & \cdots & \cdots & \beta_{2,n}\\
\beta_{1,3}^{*} & 0 & \omega_{3} & 0 & \cdots & \cdots & \vdots\\
\beta_{1,4}^{*} & \beta_{2,4}^{*} & 0 & \ddots & \ddots & \beta_{n-3,n-1} & \beta_{n-3,n}\\
\vdots & \vdots & \vdots & \ddots & \ddots & 0 & \beta_{n-2,n}\\
\beta_{1,n-1}^{*} & \vdots & \vdots & \beta_{n-3,n-1}^{*} & 0 & \omega_{n-1} & 0\\
\beta_{1,n}^{*} & \beta_{2,n}^{*} & \cdots & \beta_{n-3,n}^{*} & \beta_{n-2,n}^{*} & 0 & \omega_{n}
\end{array}\right]
\end{equation}
Again, using the quantum brachistochrone, we find that the Hamiltonian
matrix, which defines a continuous symmetry group, is constant. One
more infinite dimensional system has system Hamiltonian and constraint:

\begin{equation}
\tilde{H}[t]=\left[\begin{array}{cccccc}
\omega_{1} & 0 & 0 & \cdots & 0 & 0\\
0 & \omega_{2} & 0 & \ddots & 0 & 0\\
0 & 0 & \ddots & \ddots & \ddots & \vdots\\
\vdots & \ddots & \ddots & \ddots & 0 & 0\\
0 & 0 & \ddots & 0 & \omega_{n-1} & 0\\
0 & 0 & \cdots & 0 & 0 & \omega_{n}
\end{array}\right]
\end{equation}
\begin{equation}
\tilde{F}[t]=\left[\begin{array}{cccccc}
0 & \eta_{1,2} & \cdots & \cdots & \cdots & \eta_{1,n}\\
\eta_{1,2}^{*} & 0 & \ddots & \ddots & \ddots & \vdots\\
\vdots & \ddots & 0 & \ddots & \ddots & \vdots\\
\vdots & \ddots & \ddots & \ddots & \ddots & \vdots\\
\vdots & \ddots & \ddots & \ddots & 0 & \eta_{n-1,n}\\
\eta_{1,n}^{*} & \cdots & \cdots & \cdots & \eta_{n-1,n}^{*} & 0
\end{array}\right]
\end{equation}
\begin{equation}
\sum_{j}\omega_{j}(t)=\omega_{1}+\omega_{2}+....+\omega_{j}+....+\omega_{n-1}+\omega_{n}=0
\end{equation}
Obviously, by construction, the Hamiltonian commutes with the constraint,
and is therefore constant. The difficulty with these problems on arbitrary
dimensions is that there is no obvious way in which the Lagrange multipliers,
representing total upper bounds on the energy, generalises to a matrix
of infinite dimensions. One would hope that the sums would go over
to integrals in the correct fashion, but this remains to shown. Of
course it is possible to create these type of general Hamiltonian-constraint
systems with the geodesic property, but it would be desirable to have
expressions for field densities for matrices of infinite dimension,
which represent the continuous degrees of freedom.

\noindent \begin{flushleft}
In conclusion, we have shown within this paper how one might go about
calculating these more difficult examples on finite dimensional systems,
as well as demonstrating the links between the matrix and vector dynamic
calculus. This paper has demonstrated that the use of finite matrix
methods and algebraic techniques may yield dividends, even when the
situation is graphically complex. We have used the quantum brachistochrone
equation to derive a number of new results on SU(3); these methods
may be extended to higher-dimensional matrix groups in order to find
time-optimal flows of quantum states. The results derived are intriguing,
especially the links between the various branches of mathematics.
Our principal results may be applied within the laboratory to achieve
fast data transfer; as the method of calculation is not reliant on
adiabatic transfer or the rotating wave approximation it is theoretically
the most general and efficient scheme that may be applied. 
\par\end{flushleft}

\bibliographystyle{alpha}  \bibliographystyle{alpha}
\bibliography{abbrv,textbooks,QuantumComputation,stallisEntropy}

\end{document}